\documentclass{lmcs}
\usepackage{cite}
\usepackage{amsmath}
\usepackage{amssymb,stmaryrd,mathtools,xspace,url,bbm}
\usepackage[matrix,arrow,cmtip,curve]{xy}
\SelectTips{cm}{}\SilentMatrices\CompileMatrices
\usepackage{tikz}
\usetikzlibrary{arrows,automata,shapes}
\newcommand*\uli[1]{}

\newcommand*\cat[1]{\text{\textup{\textsf{#1}}}}
\newcommand*\HDA{\cat{HDA}}
\newcommand*\HDP{\cat{HDP}}
\newcommand*\HDT{\cat{HDT}}
\newcommand*\HDAt{\cat{HDO}}
\newcommand*\HDAh{\HDA_\textup{\textsf{h}}}
\newcommand*\HDPh{\HDP_\textup{\textsf{h}}}
\newcommand*\HDTh{\HDT_\textup{\textsf{h}}}
\newcommand*\LHDA{\cat{L}\HDA}
\newcommand*\pCub{\cat{pCub}}
\newcommand*\Cub{\cat{Cub}}
\newcommand*\LHDAh{\LHDA_\textup{\textsf{h}}}
\newcommand*\Uh{U_\textup{\textsf{h}}}
\newcommand*\Jh{J_\textup{\textsf{h}}}
\newcommand*\ttilde[1]{\tilde{ \tilde #1}}
\newcommand{\Nat}{\mathbbm{N}}

\DeclareMathOperator{\id}{id}
\newcommand*\tto[1]{\xrightarrow{#1}}
\newcommand*\tfrom[1]{\xleftarrow{#1}}
\newcommand*\from{\leftarrow}
\newcommand*\bang{\mathord{!}}

\newcommand*\ie{\textit{i.e.}\xspace}

\newcommand*\cf{\textit{cf.}\xspace}

\begin{document}

\title{Homotopy Bisimilarity for Higher-Dimensional Automata}

\author{Uli Fahrenberg \and Axel Legay}

\address{IRISA / Inria Rennes, Campus de Beaulieu, 35042 Rennes Cedex,
  France}

\email{ulrich.fahrenberg@inria.fr}

% \thanks{}
% \titlecomment{OPTIONAL comment concerning the title, \eg, if a variant
% or an extended abstract of the paper has appeared elsewehere}

\begin{abstract}
  We introduce a new category of higher-dimensional automata in which
  the morphisms are functional homotopy simulations, \ie functional
  simulations up to concurrency of independent events.  For this, we use
  unfoldings of higher-dimensional automata into higher-dimensional
  trees.  Using a notion of open maps in this category, we define
  homotopy bisimilarity.  We show that homotopy bisimilarity is
  equivalent to a straight-forward generalization of standard
  bisimilarity to higher dimensions, and that it is finer than split
  bisimilarity and incomparable with history-preserving bisimilarity.
\end{abstract}

\maketitle

\section{Introduction}

The dominant notion for behavioral equivalence of processes is
\emph{bisimulation} as introduced by Park~\cite{DBLP:conf/tcs/Park81}
and Milner~\cite{book/Milner89}.  It is compelling because it enjoys
good algebraic properties, admits several easy characterizations using
modal logics, fixed points, or game theory, and generally has low
computational complexity.

Bisimulation, or rather its underlying semantic model of
\emph{transition systems}, applies to a setting in which concurrency of
actions is the same as non-deterministic interleaving; using CCS
notation~\cite{book/Milner89}, $a| b= a. b+ b. a$.  For some
applications however, a distinction between these two is necessary,
which has led to development of so-called \emph{non-interleaving} or
\emph{truly concurrent} models such as Petri nets~\cite{book/Petri62},
event structures~\cite{DBLP:journals/tcs/NielsenPW81}, asynchronous
transition systems~\cite{Bednarczyk87-async,DBLP:journals/cj/Shields85}
and others; see~\cite{WinskelN95-Models} for a survey.

\emph{Higher-dimensional automata} (or \emph{HDA}) is another
non-interleaving formalism for reasoning about behavior of concurrent
systems.  Introduced by Pratt~\cite{Pratt91-geometry} and
van~Glabbeek~\cite{Glabbeek91-hda} in 1991 for the purpose of a
\emph{geometric} interpretation to the theory of concurrency, it has
since been shown by van~Glab\-beek~\cite{DBLP:journals/tcs/Glabbeek06}
that HDA provide a generalization (up to \emph{history-preserving
  bisimilarity}) to ``the main models of concurrency proposed in the
literature''~\cite{DBLP:journals/tcs/Glabbeek06}, including the ones
mentioned above.  Hence HDA are useful as a tool for comparing and
relating different models, and also as a modeling formalism by
themselves.

HDA are geometric in the sense that they are very similar to the
\emph{simplicial complexes} used in algebraic topology, and research on
HDA has drawn on a lot of tools and methods from geometry and algebraic
topology such as
homotopy~\cite{DBLP:journals/tcs/FajstrupRG06,DBLP:journals/mscs/Gaucher00},
homology~\cite{DBLP:conf/concur/GoubaultJ92,journals/ctopgd/Gaucher02},
and model
categories~\cite{journals/tac/Gaucher11,journals/tac/Gaucher09}, see
also the survey~\cite{DBLP:journals/mscs/Goubault00a}.

There are a number of popular notions of equivalence for HDA and other
non-interleaving models, see~\cite{DBLP:journals/acta/GlabbeekG01,
  DBLP:journals/tcs/Glabbeek06}.  \emph{Split bisimilarity} takes
interleavings of beginning and ending actions into account;
\emph{ST-bisimilarity} additionally distinguishes between different
occurrences of the same action; \emph{history-preserving bisimilarity}
takes entire computing histories into account; and \emph{hereditary
  history-preserving bisimilarity} additionally distinguishes different
possible futures of past computations.

We have in earlier work~\cite{Fahrenberg05-hda} introduced a new such
equivalence, higher-dimensional bisimilarity.  Contrary to the
previously mentioned ones, this is not a relation between
\emph{computations}, but directly at the level of states, transitions
etc.  Using \emph{unfoldings} of HDA, which geometrically are similar to
\emph{universal coverings}, we show in the present paper that this
notion is equivalent to another one, \emph{homotopy bisimilarity}, which
compares homotopy classes of computations.  Placing homotopy
bisimulation on the spectrum of non-interleaving equivalences, we show
that homotopy bisimilarity is finer than split bisimilarity and
incomparable with history-preserving bisimilarity.

% One of the most popular notions of equivalence for non-interleaving
% systems is \emph{history-preserving bisimilarity} (or
% \emph{hp-bisimilarity} for short).  It was introduced independently by
% Degano, De Nicola and Montanari in~\cite{DeganoNM89} and by Rabinovich
% and Trakhtenbrot~\cite{journals/fundinf/RabinovichT88} and then for
% event structures by van~Glabbeek and Goltz
% in~\cite{DBLP:conf/mfcs/GlabbeekG89}\uli{Replace citation!} and for
% Petri nets by Best \etal~in~\cite{DBLP:journals/acta/BestDKP91}.  One
% reason for its popularity is that it is a congruence under action
% refinement~\cite{DBLP:conf/mfcs/GlabbeekG89,DBLP:journals/acta/BestDKP91},
% another its good decidability properties: it has been shown to be
% decidable for safe Petri nets by Montanari and
% Pistore~\cite{DBLP:conf/stacs/MontanariP97}.  As a contrast, its
% cousin \emph{hereditary} hp-bisimilarity is shown undecidable for
% $1$-safe Petri nets by Jurdzi{\'n}ski, Nielsen and Srba
% in~\cite{DBLP:journals/iandc/JurdzinskiNS03}.

% In this paper we give a geometric interpretation to hp-bisimilarity for
% HDA, using the open-maps approach introduced by Joyal, Nielsen and
% Winskel in~\cite{DBLP:journals/iandc/JoyalNW96} and results from a
% previous paper~\cite{Fahrenberg05-hda} by the first author.  Using this
% interpretation, we show that hp-bisimilarity for HDA has a
% characterization directly in terms of (higher-dimensional)
% \emph{transitions} of the HDA, rather than in terms of runs as \eg~for
% Petri nets~\cite{DBLP:conf/mfcs/FroschleH99}.

Our results imply \emph{decidability} of homotopy bisimilarity for
finite HDA.  They also put homotopy bisimilarity firmly into the
open-maps framework of~\cite{DBLP:journals/iandc/JoyalNW96} and tighten
the connections between bisimilarity and weak topological
\emph{fibrations}~\cite{AdamekHRT02-weak,KurzR05-weak}.

\subsection*{Outline}

We start by reviewing the category $\HDA$ of higher-dimensional automata
introduced in~\cite{Goubault02-cmcim} in Section~\ref{se:hda}.  This is
the category used in~\cite{Fahrenberg05-hda} as a framework to define
composition, following~\cite{WinskelN95-Models}, and a notion of
bisimilarity via open maps,
following~\cite{DBLP:journals/iandc/JoyalNW96}, for HDA.  This latter
construction, together with its notion of path category, we recall in
Section~\ref{se:bisim}.

Computations in HDA are modeled by cube paths, the higher-dimensional
analogue of paths in transition systems.  These come with a notion of
homotopy which we introduce in Section~\ref{se:hdp}.  Based on homotopy
classes of cube paths we can then define the construction at the heart
of this paper, the unfolding of a HDA.

In Section~\ref{se:hdah} we introduce the category $\HDAh$ of
higher-dimensional automata up to homotopy, based on unfoldings.  We
also show in this section that unfolding provides a coreflection between
HDA and higher-dimensional trees, and between HDA-up-to-homotopy and
higher-dimensional trees.  In Section~\ref{se:open-hdah} we define
homotopy bisimilarity via open maps in $\HDAh$ and show that this is the
same as bisimilarity in $\HDA$.

All these first sections deal with unlabeled higher-dimensional
automata.  In Section~\ref{se:labels}, we introduce labels using an
arrow category construction and show that things can easily be
transferred to the labeled setting.  In Section~\ref{se:comparison} we
compare homotopy bisimilarity to other equivalence notions for
non-interleaving models.

\subsection*{Acknowledgements}

The authors wish to thank Rob van Glabbeek for many useful discussions
on the subject of this paper, and the organizers of SMC 2014 in Lyon for
providing a forum for these discussions.

\section{Higher-Dimensional Automata}
\label{se:hda}

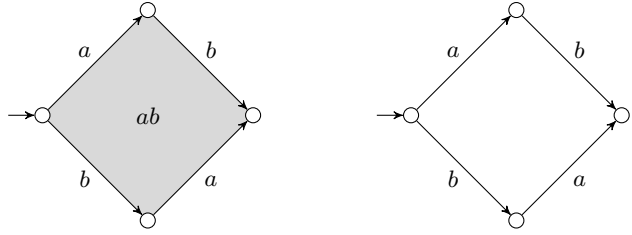
\begin{figure}
  \centering
  \begin{tikzpicture}[->,>=stealth',auto,scale=.7]
    \tikzstyle{every node}=[font=\footnotesize,initial text=]
    \tikzstyle{every state}=[fill=white,shape=circle,inner
    sep=.5mm,minimum size=2mm]
    \begin{scope}
      \path[fill=black!15] (0,0) to (2,2) to (4,0) to (2,-2);
      \node[state,initial] (s0) at (0,0) {};
      \node[state] (s1) at (2,2) {};
      \node[state] (s2) at (2,-2) {};
      \node[state] (s3) at (4,0) {};
      \path (s0) edge (s1);
      \path (s0) edge (s2);
      \path (s1) edge (s3);
      \path (s2) edge (s3);
      \node at (.8,1.2) {$a$};
      \node at (.8,-1.2) {$b$};
      \node at (3.2,1.25) {$b$};
      \node at (3.2,-1.25) {$a$};
      \node at (2,0) {$ab$};
    \end{scope}
    \begin{scope}[xshift=7cm]
      \node[state,initial] (s0) at (0,0) {};
      \node[state] (s1) at (2,2) {};
      \node[state] (s2) at (2,-2) {};
      \node[state] (s3) at (4,0) {};
      \path (s0) edge (s1);
      \path (s0) edge (s2);
      \path (s1) edge (s3);
      \path (s2) edge (s3);
      \node at (.8,1.2) {$a$};
      \node at (.8,-1.2) {$b$};
      \node at (3.2,1.25) {$b$};
      \node at (3.2,-1.25) {$a$};
    \end{scope}
  \end{tikzpicture}
  \caption{%
    \label{fi:independence}
    HDA for the CCS expressions $a| b$ (left) and $a. b+ b. a$ (right).
    In the left HDA, the square is filled in by a two-dimensional
    transition labeled $ab$, signifying independence of events $a$ and
    $b$.  On the right, $a$ and $b$ are not independent.%
  }
\end{figure}

As a formalism for concurrent behavior, HDA have the specific feature
that they can express all higher-order dependencies between events in a
concurrent system.  Like for transition systems, they consist of states
and transitions which are labeled with events.  Now if two transitions
from a state, with labels $a$ and $b$ for example, are independent, then
this is expressed by the existence of a \emph{two-dimensional}
transition with label $ab$.  Fig.~\ref{fi:independence} shows two
examples; on the left, transitions $a$ and $b$ are independent, on the
right, they can merely be executed in any order.  Hence for HDA, as
indeed for any formalism employing the so-called \emph{true concurrency}
paradigm, the algebraic law $a| b= a. b+ b. a$ does \emph{not} hold;
concurrency is not the same as interleaving.

The above considerations can equally be applied to sets of more than two
events: if three events $a$, $b$, $c$ are independent, then this is
expressed using a three-dimensional transition labeled $abc$.  Hence
this is different from mutual pairwise independence (expressed by
transitions $ab$, $ac$, $bc$), a distinction which cannot be made in
formalisms such as asynchronous transition
systems~\cite{Bednarczyk87-async,DBLP:journals/cj/Shields85} or
transition systems with independence~\cite{WinskelN95-Models} which only
consider binary independence relations.

An unlabeled HDA is essentially a pointed precubical set as defined
below.  For labeled HDA, one can pass to an arrow category; this is what
we shall do in Section~\ref{se:labels}.  Until then, we concentrate on
the unlabeled case.

\begin{figure}
  \centering
  \begin{tikzpicture}[auto,scale=1]
    \tikzstyle{every node}=[font=\footnotesize]
    \tikzstyle{every state}=[fill=white,shape=circle,inner
    sep=.5mm,minimum size=2mm]
    \path[fill=black!15] (0,0) to (2,0) to (2,2) to (0,2) to (0,0);
    \path (0,0) edge (2,0);
    \path (2,0) edge (2,2);
    \path (0,0) edge (0,2);
    \path (0,2) edge (2,2);
    \node[state] at (0,0) {};
    \node[state] at (2,0) {};
    \node[state] at (0,2) {};
    \node[state] at (2,2) {};
    \node at (1,1) {$x$};
    \node at (-.4,1.05) {$\delta_1^0 x$};
    \node at (2.4,1.05) {$\delta_1^1 x$};
    \node at (1,-.25) {$\delta_2^0 x$};
    \node at (1,2.25) {$\delta_2^1 x$};
    \node at (-1,-.35) {$\delta_1^0 \delta_2^0 x= \delta_1^0
      \delta_1^0 x$};
    \node at (-1,2.35) {$\delta_1^0 \delta_2^1 x= \delta_1^1
      \delta_1^0 x$};
    \node at (3,-.35) {$\delta_1^1 \delta_2^0 x= \delta_1^0
      \delta_1^1 x$};
    \node at (3,2.35) {$\delta_1^1 \delta_2^1 x= \delta_1^1
      \delta_1^1 x$};
  \end{tikzpicture}
  \caption{%
    \label{fi:2cubefaces}
    A $2$-cube $x$ with its four faces $\delta_1^0 x$, $\delta_1^1 x$,
    $\delta_2^0 x$, $\delta_2^1 x$ and four corners.
  }
\end{figure}
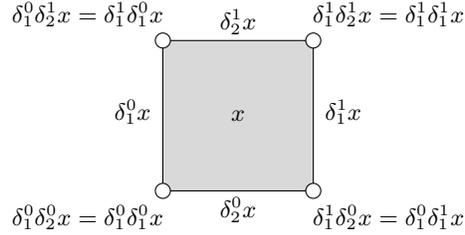

A \emph{precubical set} is a graded set $X= \{ X_n\}_{ n\in \Nat}$
together with mappings $\delta_k^\nu:X_n\to X_{ n- 1}$, $k\in\{ 1,\dots,
n\}$, $\nu\in\{ 0, 1\}$, satisfying the \emph{precubical identity}
\begin{equation}
  \label{eq:pcub}
  \delta_k^\nu \delta_\ell^\mu= \delta_{ \ell- 1}^\mu
  \delta_k^\nu \qquad( k< \ell)\,.
\end{equation}
The mappings $\delta_k^\nu$ are called \emph{face maps}, and elements of
$X_n$ are called \emph{$n$-cubes}.  As above, we shall usually omit the
extra subscript $(n)$ in the face maps.  Faces $\delta_k^0 x$ of an
element $x\in X$ are to be thought of as \emph{lower faces}, $\delta_k^1
x$ as \emph{upper faces}.  The precubical identity expresses the fact
that $( n- 1)$-faces of an $n$-cube meet in common $( n- 2)$-faces, see
Fig.~\ref{fi:2cubefaces} for an example of a $2$-cube and its faces.

We will always assume the sets $X_n$ to be disjoint.  For an $n$-cube
$x\in X_n$, we denote by $\dim x= n$ its \emph{dimension}.

\emph{Morphisms} $f: X\to Y$ of precubical sets are graded mappings
$f=\{ f_n: X_n\to Y_n\}_{ n\in \Nat}$ which commute with the face maps:
$\delta_k^\nu\circ f_n= f_{ n- 1}\circ \delta_k^\nu$ for all $n\in
\Nat$, $k\in\{ 1,\dots, n\}$, $\nu\in\{ 0, 1\}$.  This defines a
category $\pCub$ of precubical sets and morphisms.

It can be shown~\cite{GrandisM03-Site} that the category $\pCub$ is
complete and cocomplete, with point-wise limits and colimits.  In
elementary terms this means that, for instance, the \emph{product} $Z=
X\times Y$of two precubical sets $X$, $Y$ is given by $Z_n= X_n\times
Y_n$ and face maps $\delta_k^\nu( x, y)=( \delta_k^\nu x, \delta_k^\nu
y)$.  Likewise, a \emph{precubical subset} $Y\subseteq X$ of $X\in
\pCub$ is a precubical set $Y$ for which $Y_n\subseteq X_n$ for all
$n$.

A \emph{pointed} precubical set is a precubical set $X$ with a specified
$0$-cube $i\in X_0$, and a pointed morphism is one which respects the
point.  This defines a category which is isomorphic to the comma
category $*\downarrow \pCub$, where $*\in \pCub$ is the precubical set
with one $0$-cube and no other $n$-cubes.  Note that $*$ is \emph{not}
terminal in $\pCub$ (instead, the terminal object is the somewhat
unwieldy infinite-dimensional precubical set with one cube in every
dimension).

\begin{defi}
  \label{de:hda}
  The category of \emph{higher-dimensional automata} is the comma
  category $\HDA= *\downarrow \pCub$, with objects pointed
  precubical sets and morphisms commutative diagrams
  \begin{equation*}
    \xymatrix@C=1.3em@R=1.1em{%
      & {*} \ar[dl] \ar[dr] \\ X \ar[rr]_f && Y\,.
    }
  \end{equation*}
\end{defi}

Hence a one-dimensional HDA is a transition system; indeed, the category
of transition systems~\cite{WinskelN95-Models} is isomorphic to the full
subcategory of $\HDA$ spanned by the one-dimensional objects.  Similarly
one can show~\cite{Goubault02-cmcim} that the category of asynchronous
transition systems is isomorphic to the full subcategory of $\HDA$
spanned by the (at most) two-dimensional objects.  The category $\HDA$
as defined above was used in~\cite{Fahrenberg05-hda} to provide a
categorical framework (in the spirit of~\cite{WinskelN95-Models}) for
parallel composition of HDA.  In this article we also introduced a
notion of higher-dimensional bisimilarity which we will review in the
next section.

\section{Path Objects, Open Maps and Bisimilarity}
\label{se:bisim}

With the purpose of introducing bisimilarity via \emph{open maps} in the
sense of~\cite{DBLP:journals/iandc/JoyalNW96}, we identify here a
subcategory of $\HDA$ consisting of path objects and path-extending
morphisms.  We say that a precubical set $X$ is a \emph{precubical path
  object} if there is a (necessarily unique) sequence $( x_1,\dots,
x_m)$ of elements in $X$ such that $x_i\ne x_j$ for $i\ne j$,
\begin{itemize}
\item for each $x\in X$ there is $j\in\{ 1,\dots, m\}$ for which
  $\smash{ x= \delta_{ k_1}^{ \nu_1}\cdots \delta_{ k_p}^{ \nu_p} x_j}$
  for some indices $\nu_1,\dots, \nu_p$ and a \emph{unique} sequence
  $k_1<\dots< k_p$, and
\item for each $j= 1,\dots, m- 1$, there is $k\in \Nat$ for which $x_j=
  \delta_k^0 x_{ j+ 1}$ or $x_{ j+ 1}= \delta_k^1 x_j$.
\end{itemize}

Note that precubical path objects are \emph{non-selflinked} in the sense
of~\cite{DBLP:journals/tcs/FajstrupRG06}.  If $X$ and $Y$ are precubical
path objects with representations $( x_1,\dots, x_m)$, $( y_1,\dots,
y_p)$, then a morphism $f: X\to Y$ is called a \emph{cube path
  extension} if $x_j= y_j$ for all $j= 1,\dots, m$ (hence $m\le p$).

\begin{defi}
  The category $\HDP$ of \emph{higher-dimensional paths} is the
  subcategory of $\HDA$ which as objects has pointed precubical paths,
  and whose morphisms are generated by pointed cube path extensions and
  isomorphisms.
\end{defi}

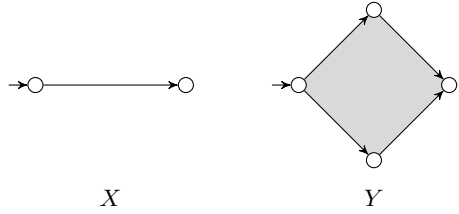
\begin{figure}
  \centering
  \begin{tikzpicture}[->,>=stealth',auto,scale=.5]
    \tikzstyle{every node}=[font=\footnotesize,initial text=]
    \tikzstyle{every state}=[fill=white,shape=circle,inner
    sep=.5mm,minimum size=2mm]
    \begin{scope}
      \node[state,initial] (0) at (0,0) {};
      \node[state] (1) at (4,0) {};
      \path (0) edge (1);
      \node at (2,-3) {$X$};
    \end{scope}
    \begin{scope}[xshift=7cm]
      \path[fill=black!15] (0,0) to (2,2) to (4,0) to (2,-2);
      \node[state,initial] (0) at (0,0) {};
      \node[state] (1) at (2,2) {};
      \node[state] (2) at (2,-2) {};
      \node[state] (3) at (4,0) {};
      \path (0) edge (1);
      \path (1) edge (3);
      \path (0) edge (2);
      \path (2) edge (3);
      \node at (2,-3) {$Y$};
    \end{scope}
  \end{tikzpicture}
  \caption{%
    \label{fi:hdpnotfull}
    Two higher-dimensional paths with no $\HDP$-morphism between them.
  }
\end{figure}

\begin{exa}
  $\HDP$ is not a full subcategory of $\HDA$: If $X$ and $Y$ are the two
  higher-dimensional paths depicted in Fig.~\ref{fi:hdpnotfull}, then
  none of the two mappings $X\to Y$ is a $\HDP$-morphism.
\end{exa}

A \emph{cube path} in a precubical set $X$ is a morphism $P\to X$ from a
precubical path object $P$.  In elementary terms, this is a sequence $(
x_1,\dots, x_m)$ of elements of $X$ such that for each $j= 1,\dots, m-
1$, there is $k\in \Nat$ for which $x_j= \delta_k^0 x_{ j+ 1}$ (start of
a new part of a computation) or $x_{ j+ 1}= \delta_k^1 x_j$ (end of a
computation part).

\begin{figure}
  \centering
  \begin{tikzpicture}[->,>=stealth',auto,scale=.9]
    \tikzstyle{every node}=[font=\footnotesize,initial text=]
    \tikzstyle{every state}=[fill=white,shape=circle,inner
    sep=.5mm,minimum size=3mm]
    \path[fill=black!15] (2,0) to (4,0) to (4,2) to (2,2);
    \node[state,initial] (i) at (0,0) {};
    \node[state] (x) at (2,0) {};
    \node[state] (z) at (4,2) {};
    \path (i) edge (x); %a
    \path (x) edge (4,0); %b
    \path (4,0) edge (z); %c
    \path (z) edge (6,2); %d
    \node at (0,-.35) {$i$};
    \node at (1,-.25) {$a$};
    \node at (2,-.4) {$x$};
    \node at (3,-.2) {$b$};
    \node at (3,1) {$bc$};
    \node at (4.2,.97) {$c$};
    \node at (4,2.3) {$z$};
    \node at (5,2.2) {$d$};
  \end{tikzpicture}
  \caption{%
    \label{fi:cubepath}
    Graphical representation of the two-dimensional cube path $( i, a,
    x, b, bc, c, z, d)$.  Its computational interpretation is that $a$
    is executed first, then execution of $b$ starts, and while $b$ is
    running, $c$ starts to execute.  After this, $b$ finishes, then $c$,
    and then execution of $d$ is started.  Note that the computation is
    partial, as $d$ does not finish.
  }
\end{figure}
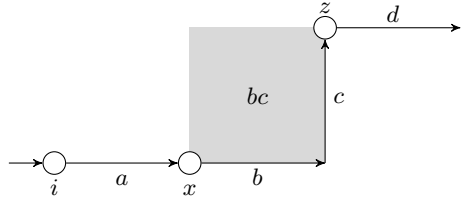

Cube paths were introduced in~\cite{Glabbeek91-hda}, where they are
simply called paths.  They are intended to model (partial) computations
of HDA.  We show an example of a cube path in Fig.~\ref{fi:cubepath}.

A cube path in a HDA $i: *\to X$ is \emph{pointed} if $x_1= i$, hence if
it is a pointed morphism $P\to X$ from a higher-dimensional path $P$.
We will say that a cube path $( x_1,\dots, x_m)$ is \emph{from} $x_1$
\emph{to} $x_m$, and that a cube $x\in X$ in a HDA $X$ is
\emph{reachable} if there is a pointed cube path to $x$ in $X$.

Cube paths can be \emph{concatenated} if the end of one is compatible
with the beginning of the other: If $\rho=( x_1,\dots, x_m)$ and
$\sigma=( y_1,\dots, y_p)$ are cube paths with $y_1= \delta_k^1 x_m$ or
$x_m= \delta_k^0 y_1$ for some $k$, then their \emph{concatenation} is
the cube path $\rho* \sigma=( x_1,\dots, x_m, y_1,\dots, y_p)$.  We say
that $\rho$ is a \emph{prefix} of $\chi$ and write $\rho\sqsubseteq
\chi$ if there is a cube path $\rho$ for which $\chi=\rho* \sigma$.

\begin{defi}
  \label{de:open_hda}
  A pointed morphism $f: X\to Y$ in $\HDA$ is an \emph{open map} if it
  has the right lifting property with respect to $\HDP$, \ie~if it is
  the case that there is a lift $r$ in any commutative diagram as below,
  for morphisms $g: P\to Q\in \HDP$, $p: P\to X, q: Q\to Y\in \HDA$:
  \begin{equation*}
    \xymatrix{%
      P \ar[r]^p \ar[d]_g & X \ar[d]^f \\ Q \ar[r]_q \ar@{.>}[ur]|r &
      Y
    }
  \end{equation*}
  HDA $X$, $Y$ are \emph{hd-bisimilar} if there is $Z\in \HDA$ and a
  span of open maps $X\from Z\to Y$ in $\HDA$.
\end{defi}

It follows straight from the definition that composites of open maps again
are open.  By the next lemma, morphisms are open precisely when they
have a zig-zag property similar to the one
of~\cite{DBLP:journals/iandc/JoyalNW96}.

\begin{lem}
  \label{le:open}
  For a morphism $f: X\to Y\in \HDA$, the following are equivalent:
  \begin{enumerate}
  \item\label{enu:open.box} $f$ is open;
  \item\label{enu:open.onestep} for any reachable $x_1\in X$ and any
    $y_2\in Y$ with $f( x_1)= \delta_k^0 y_2$ for some $k$, there is
    $x_2\in X$ for which $x_1= \delta_k^0 x_2$ and $y_2= f( x_2)$;
  \item\label{enu:open.cubepath} for any reachable $x_1\in X$ and any
    cube path $( y_1,\dots, y_m)$ in $Y$ with $y_1= f( x_1)$, there is a
    cube path $( x_1,\dots, x_m)$ in $X$ for which $y_j= f( x_j)$ for
    all $j= 1,\dots, m$.
  \end{enumerate}
\end{lem}

\proof%
  For the implication
  \eqref{enu:open.box}~$\Longrightarrow$~\eqref{enu:open.onestep}, let
  $p: P\to X$ be a pointed cube path with $P$ represented by $(
  p_1,\dots, p_m)$ and $p( p_m)= x_1$.  Let $p_{ m+ 1}$ be a cube of
  dimension one higher than $p_m$, set $p_m= \delta_k^0 p_{ m+ 1}$, and
  let $Q$ be the higher-dimensional path represented by $( p_1,\dots,
  p_m, p_{ m+ 1})$.  Let $g: P\to Q$ be the inclusion, and define $q:
  Q\to Y$ by $q( p_j)= f( p( p_j))$ for $j= 1,\dots, m$ and $q( p_{ m+
    1})= y_2$.  We have a lift $r: Q\to X$ and can set $x_2= r( p_{ m+
    1})$.

  The implication
  \eqref{enu:open.onestep}~$\Longrightarrow$~\eqref{enu:open.cubepath}
  can be easily shown by induction.  The case $y_m= \delta_k^0 y_{ m+
    1}$ follows directly from~\eqref{enu:open.onestep}, and the case
  $y_{ m+ 1}= \delta_k^1 y_m$ is clear by $\delta_k^1 \circ f= f\circ
  \delta_k^1$.

  To finish the proof, we show the implication
  \eqref{enu:open.cubepath}~$\Longrightarrow$~\eqref{enu:open.box}.  Let
  \begin{equation*}
    \xymatrix{%
      P \ar[r]^p \ar[d]_g & X \ar[d]^f \\ Q \ar[r]_q & Y
    }
  \end{equation*}
  be a commutative diagram, with $P$ represented by $( p_1,\dots, p_m)$.
  Up to isomorphism we can assume that $Q$ is represented by $(
  p_1,\dots, p_m, p_{ m+ 1},\dots, p_t)$ and that $g$ is the inclusion.
  The cube $p( p_m)$ is reachable in $X$, and $( q( p_m),\dots, q(
  p_t))$ is a cube path in $Y$ which starts in $q( p_m)= f( p( p_m))$.
  Hence we have a cube path $( x_m,\dots, x_t)$ in $X$ with $x_m= p(
  p_m)$ and $q( p_j)= f( x_j)$ for all $j= m,\dots, t$, and we can
  define a lift $r: Q\to X$ by $r( p_j)= p( p_j)$ for $j= 1,\dots, m$
  and $r( p_j)= x_j$ for $j= m+ 1,\dots, t$.
\qed

\begin{thm}
  \label{th:bisim}
  For HDA $i: *\to X$, $j: * \to Y$, the following are equivalent:
  \begin{enumerate}
  \item\label{enu:bisim.box} $X$ and $Y$ are hd-bisimilar;
  \item\label{enu:bisim.onestep} there exists a precubical subset
    $R\subseteq X\times Y$ for which $( i, j)\in R$, and such that for
    all 
    % reachable $x_1\in X$, $y_1\in Y$ with
    $( x_1, y_1)\in R$,
    \begin{itemize}
    \item for any $x_2\in X$ for which $x_1= \delta_k^0 x_2$ for some
      $k$, there exists $y_2\in Y$ for which $y_1= \delta_k^0 y_2$ and
      $( x_2, y_2)\in R$,
    \item for any $y_2\in Y$ for which $y_1= \delta_k^0 y_2$ for some
      $k$, there exists $x_2\in X$ for which $x_1= \delta_k^0 x_2$ and
      $( x_2, y_2)\in R$;
    \end{itemize}
  \item\label{enu:bisim.cubepath} there exists a precubical subset
    $R\subseteq X\times Y$ for which $( i, j)\in R$, and such that for
    all 
    % reachable $x_1\in X$, $y_1\in Y$ with
    $( x_1, y_1)\in R$,
    \begin{itemize}
    \item for any cube path $( x_1,\dots, x_m)$ in $X$, there exists a
      cube path $( y_1,\dots, y_m)$ in $Y$ with $( x_p, y_p)\in R$ for
      all $p= 1,\dots, m$,
    \item for any cube path $( y_1,\dots, y_m)$ in $Y$, there exists a
      cube path $( x_1,\dots, x_m)$ in $X$ with $( x_p, y_p)\in R$ for
      all $p= 1,\dots, m$.
    \end{itemize}
  \end{enumerate}
\end{thm}

% Note that the requirement that $R$ be a precubical subset, in
% items~\eqref{enu:bisim.onestep} and~\eqref{enu:bisim.cubepath} above,
% is equivalent to saying that whenever $( x, y)\in R$, then also $(
% \delta_k^\nu x, \delta_k^\nu y)\in R$ for any $k$ and $\nu\in\{ 0,
% 1\}$.

\proof%
  For the implication
  \eqref{enu:bisim.box}~$\Longrightarrow$~\eqref{enu:bisim.onestep}, let
  $X\tfrom f Z\tto g Y$ be a span of open maps and define $R=\{( x,
  y)\in X\times Y\mid \exists z\in Z: x= f( z), y= g( z)\}$.  Then $( i,
  j)\in R$ because $f$ and $g$ are pointed morphisms, and the other
  properties follow by Lemma~\ref{le:open}.  The implication
  \eqref{enu:bisim.onestep}~$\Longrightarrow$~\eqref{enu:bisim.cubepath}
  can be shown by a simple induction, and for the implication
  \eqref{enu:bisim.cubepath}~$\Longrightarrow$~\eqref{enu:bisim.box},
  the projections give a span $X\tfrom{ \pi_1} R\tto{ \pi_2} Y$ and are
  open by Lemma~\ref{le:open}.
\qed

\section{Homotopies and Unfoldings}
\label{se:hdp}

In order to connect our notion of hd-bisimilarity with other common
notions, we need to introduce in which cases different cube paths are
equivalent due to independence of actions.
Following~\cite{DBLP:journals/tcs/Glabbeek06}, we model this equivalence
by a combinatorial version of \emph{homotopy} which is an extension of
the equivalence defining \emph{Mazurkiewicz
  traces}~\cite{Mazurkiewicz77}.

We say that cube paths $( x_1,\dots, x_m)$, $( y_1,\dots, y_m)$ are
\emph{adjacent} if $x_1= y_1$, $x_m= y_m$, there is precisely one index
$p\in\{ 1,\dots, m\}$ at which $x_p\ne y_p$, and
\begin{itemize}
\item $x_{ p- 1}= \delta_k^0 x_p$, $x_p= \delta_\ell^0 x_{ p+ 1}$, $y_{
    p- 1}= \delta_{ \ell- 1}^0 y_p$, and $y_p= \delta_k^0 y_{ p+ 1}$ for
  some $k< \ell$, or vice versa,
\item $x_p= \delta_k^1 x_{ p- 1}$, $x_{ p+ 1}= \delta_\ell^1 x_p$, $y_p=
  \delta_{ \ell- 1}^1 y_{ p- 1}$, and $y_{ p+ 1}= \delta_k^1 y_p$ for
  some $k< \ell$, or vice versa,
\item $x_p= \delta_k^0 \delta_\ell^1 y_p$, $y_{ p- 1}= \delta_k^0 y_p$,
  and $y_{ p+ 1}= \delta_\ell^1 y_p$ for some $k< \ell$, or vice versa,
  or
\item $x_p= \delta_k^1 \delta_\ell^0 y_p$, $y_{ p- 1}= \delta_\ell^0
  y_p$, and $y_{ p+ 1}= \delta_k^1 y_p$ for some $k< \ell$, or vice
  versa.
\end{itemize}

\begin{figure}
  \centering
  \begin{tikzpicture}[->,>=stealth',auto,scale=.8]
    \tikzstyle{every node}=[font=\small,initial text=]
    \tikzstyle{every state}=[fill=white,shape=circle,inner
    sep=.5mm,minimum size=2mm]
    \begin{scope}
      \path[fill=black!15] (2,0) to (4,0) to (4,2) to (2,2);
      \node[state,initial] (i) at (2,-2) {};
      \node[state] (x) at (2,0) {};
      \node[state] (z) at (4,2) {};
      \path (i) edge (x); %a
      \path (x) edge (4,0); %b
      \path (4,0) edge (z); %c
      \path (z) edge (4,4); %d
      \node at (2.3,-2) {$i$};
      \node at (1.65,0) {$x$};
      \node at (2.2,-1) {$a$};
      \node at (3,-.25) {$b$};
      \node at (3,1) {$bc$};
      \node at (4.2,.97) {$c$};
      \node at (4.3,2) {$z$};
      \node at (4.2,3) {$d$};
    \end{scope}
    \begin{scope}[xshift=4.5cm]
      \path[fill=black!15] (2,0) to (4,0) to (4,2) to (2,2);
      \node[state,initial] (i) at (2,-2) {};
      \node[state] (x) at (2,0) {};
      \node[state] (z) at (4,2) {};
      \path (i) edge (x); %a
      \path (x) edge (2,2);
      \path (4,0) edge (z); %c
      \path (z) edge (4,4); %d
      \node at (2.3,-2) {$i$};
      \node at (1.65,0) {$x$};
      \node at (2.2,-1) {$a$};
      \node at (1.8,.97) {$c$};
      \node at (3,1) {$bc$};
      \node at (4.2,.97) {$c$};
      \node at (4.3,2) {$z$};
      \node at (4.2,3) {$d$};
    \end{scope}
    \begin{scope}[xshift=9cm]
      \path[fill=black!15] (2,0) to (4,0) to (4,2) to (2,2);
      \node[state,initial] (i) at (2,-2) {};
      \node[state] (x) at (2,0) {};
      \node[state] (z) at (4,2) {};
      \path (i) edge (x); %a
      \path (x) edge (2,2);
      \path (2,2) edge (z);
      \path (z) edge (4,4); %d
      \node at (2.3,-2) {$i$};
      \node at (1.65,0) {$x$};
      \node at (2.2,-1) {$a$};
      \node at (1.8,.97) {$c$};
      \node at (3,2.25) {$b$};
      \node at (3,1) {$bc$};
      \node at (4.3,2) {$z$};
      \node at (4.2,3) {$d$};
    \end{scope}
    \begin{scope}[xshift=13.5cm]
      \node[state,initial] (i) at (2,-2) {};
      \node[state] (x) at (2,0) {};
      \node[state] (y) at (2,2) {};
      \node[state] (z) at (4,2) {};
      \path (i) edge (x); %a
      \path (x) edge (y);
      \path (y) edge (z);
      \path (z) edge (4,4); %d
      \node at (2.3,-2) {$i$};
      \node at (1.65,0) {$x$};
      \node at (1.65,1.95) {$y$};
      \node at (2.2,-1) {$a$};
      \node at (1.8,.97) {$c$};
      \node at (3,2.25) {$b$};
      \node at (4.3,2) {$z$};
      \node at (4.2,3) {$d$};
    \end{scope}
  \end{tikzpicture}
  \caption{%
    \label{fi:homcubepath}
    Graphical representation of the cube path homotopy $( i, a, x, b, bc, c, z,
    d)\sim$ $( i, a, x, c, bc, c, z, d)\sim( i, a, x, c, bc, b, z, d)\sim(
    i, a, x, c, y, b, z, d)$.  }
\end{figure}
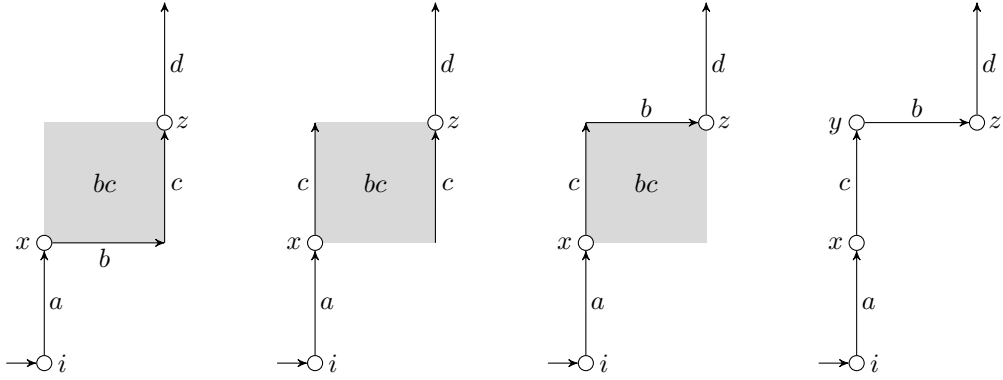

\emph{Homotopy} of cube paths is the reflexive, transitive closure of
the adjacency relation.  We denote homotopy of cube paths using the
symbol $\sim$, and the homotopy class of a cube path $( x_1,\dots, x_m)$
is denoted $[ x_1,\dots, x_m]$.  The intuition of adjacency is rather
simple, even though the combinatorics may look complicated, see
Fig.~\ref{fi:homcubepath}.  Note that adjacencies come in two basic
``flavors'': the first two above in which the dimensions of $x_\ell$ and
$y_\ell$ are the same, and the last two in which they differ by $2$.

The following lemma shows that, not surprisingly, cube paths entirely
contained in one cube are homotopic (provided that they share
endpoints).

\begin{lem}
  \label{le:hom_fullcube}
  Let $x\in X_n$ in a precubical set $X$ and $( k_1,\dots, k_n)$, $(
  \ell_1,\dots, \ell_n)$ sequences of indices with $k_j, \ell_j\le j$
  for all $j= 1,\dots, n$.  Let $x_j= \delta_{ k_j}^0\cdots \delta_{
    k_n}^0 x$, $y_j= \delta_{ \ell_j}^0\cdots \delta_{ \ell_n}^0 x$.
  Then the cube paths $( x_1,\dots, x_n, x)\sim( y_1,\dots, y_n, x)$.
\end{lem}

\proof%
    ({{\cf~\cite[Ex.~2.15]{Fajstrup05-cubcomp}}}).
  We can represent a cube path $( x_1,\dots, x_n, x)$ as above by an
  element $( p_1,\dots, p_n)$ of the symmetric group $S_n$ by setting
  $p_n= k_n$ and, working backwards, $p_j=(\{ 1,\dots, n\}\setminus\{
  p_{ j+ 1},\dots, p_n\})[ k_j]$, denoting by this the $k_j$-largest
  element of the set in parentheses.  This introduces a bijection
  between the set of cube paths from the lower left corner of $x$ to $x$
  on the one hand, and elements of $S_n$ on the other hand, and under
  this bijection adjacencies of cube paths are transpositions in $S_n$.
  These generate all of $S_n$, hence all such cube paths are
  homotopic.
\qed

We extend concatenation and prefix to homotopy classes of cube paths
by defining $[ x_1,\dots, x_m]*[ y_1,\dots, y_p]=[ x_1,\dots, x_m,
y_1,\dots, y_p]$ and saying that $\tilde x\sqsubseteq \tilde z$, for
homotopy classes $\tilde x$, $\tilde z$ of cube paths, if there are $(
x_1,\dots, x_m)\in \tilde x$ and $( z_1,\dots, z_q)\in \tilde z$ for
which $( x_1,\dots, x_m)\sqsubseteq( z_1,\dots, z_q)$.  It is easy to
see that concatenation is well-defined, and that $\tilde x\sqsubseteq
\tilde z$ if and only if there is a homotopy class $\tilde y$ for which
$\tilde z= \tilde x* \tilde y$.

Using homotopy classes of cube paths, we can now define the
\emph{unfolding} of a HDA.  Unfoldings of HDA are similar to unfoldings
of transition systems~\cite{WinskelN95-Models} or Petri
nets~\cite{DBLP:journals/tcs/NielsenPW81,DBLP:conf/fsttcs/HaymanW08},
but also to \emph{universal covering spaces} in algebraic topology.  The
intention is that the unfolding of a HDA captures all its computations,
up to homotopy.

We say that a HDA $X$ is a \emph{higher-dimensional tree} if it holds
that for any $x\in X$, there is precisely one homotopy class of pointed
cube paths to $x$.  The full subcategory of $\HDA$ spanned by the
higher-dimensional trees is denoted $\HDT$.  Note that any
higher-dimensional path is a higher-dimensional tree; indeed there is an
inclusion $\HDP\hookrightarrow \HDT$.

\begin{defi}
  \label{de:unfold}
  The \emph{unfolding} of a HDA $i: *\to X$ consists of a HDA $\tilde i:
  *\to \tilde X$ and a pointed \emph{projection} morphism $\pi_X: \tilde
  X\to X$, which are defined as follows:
  \begin{itemize}
  \item $\tilde X_n=\big\{[ x_1,\dots, x_m]\mid( x_1,\dots, x_m)$
    pointed cube path in $X, x_m\in X_n\big\}$; $\tilde i=[ i]$
  \item $\tilde \delta_k^0[ x_1,\dots, x_m]=\big\{( y_1,\dots, y_p)\mid
    y_p= \delta_k^0 x_m, ( y_1,\dots, y_p, x_m)\sim(
    x_1,\dots,x_m)\big\}$
  \item $\tilde \delta_k^1[ x_1,\dots, x_m]=[ x_1,\dots, x_m,
    \delta_k^1 x_m]$
  \item $\pi_X[ x_1,\dots, x_m]= x_m$
  \end{itemize}
\end{defi}

\begin{prop}
  \label{th:univ_cov}
  The unfolding $( \tilde X, \pi_X)$ of a HDA $X$ is well-defined, and
  $\tilde X$ is a higher-dimensional tree.  If $X$ itself is a
  higher-dimensional tree, then the projection $\pi_X: \tilde X\to X$ is
  an isomorphism.
\end{prop}

Before proving the proposition, we need an auxiliary notion of
\emph{fan-shaped} cube path together with a technical lemma.  Say that a
cube path $( x_1,\dots, x_m)$ in a precubical set $X$, with $x_m\in
X_n$, is fan-shaped if
\begin{equation*}
  x_j\in
  \begin{cases}
    X_0 &\text{for } 1\le j\le m- n\text{ odd,} \\
    X_1 &\text{for } 1\le j\le m- n\text{ even,} \\
    X_{ n+ j- m} &\text{for } m- n< j\le m\,.
  \end{cases}
\end{equation*}
Hence a fan-shaped cube path is a one-dimensional path up to the point
where it needs to build up to hit the possibly high-dimensional end cube
$x_m$; in computational terms, it is \emph{serialized}.

\begin{lem}
  \label{le:fan}
  Any pointed cube path in a higher-dimensional automaton $i: *\to X$ is
  homotopic to a fan-shaped one.
\end{lem}

\proof%
Let us first introduce some notation: For any pointed cube path $(
x_1,\dots, x_m)$, let $n_j= \dim x_j$ be the $j$-th component's
dimension, and let $T( x_1,\dots, x_m)= n_1+\cdots+ n_m$.  An easy
induction shows that $j- n_j$ is odd for all $j$.  Also, $T( x_1,\dots,
x_m)\ge \frac12( n_m^2+ m- 1)$, with equality if and only if $(
x_1,\dots, x_m)$ is fan-shaped.

Next we show that $n_1+\cdots+ n_m\equiv \frac12( n_m^2+ m- 1)\mod 2$.
By oddity of $j- n_j$ we have $\sum_{ j= 1}^m n_j- \sum_{ j= 1}^m
j\equiv m\mod 2$, and also $\frac12( n_m^2+ m- 1)- \sum_{ j= 1}^m j=
\frac12( n_m^2- m^2- 1)\equiv m\mod 2$, hence the claim follows.

We can now finish the proof by showing how to convert a cube path $(
x_1,\dots, x_m)$ with $T( x_1,\dots, x_m)> \frac12( n_m^2+ m- 1)$ into
an adjacent cube path $( x_1',\dots, x_m')$ which has $T( x_1',\dots,
x_m')= T( x_1,\dots, x_m)- 2$, essentially by replacing one of its
cubes, called $x_\ell$ below, with another one of dimension $n_\ell- 2$.

If $( x_1,\dots, x_m)$ is a cube path which is not fan-shaped, then
there is an index $\ell\in\{3,\dots, m- 1\}$ for which $n_\ell\ge 2$,
$x_{ \ell- 1}= \delta_{ k_2}^0 x_\ell$ for some $k_2$, and $x_{ \ell+
  1}= \delta_{ k_3}^1 x_\ell$ for some $k_3$.  Assuming $\ell$ to be the
\emph{least} such index, we must also have $x_{ \ell- 2}= \delta_{
  k_1}^0 x_{ \ell- 1}$ for some $k_1$.

Now if $k_2< k_3$, then $\delta_{ k_2}^0 x_{ \ell+ 1}= \delta_{ k_2}^0
\delta_{ k_3}^1 x_\ell= \delta_{ k_3- 1}^1 \delta_{ k_2}^0 x_\ell=
\delta_{ k_3- 1}^1 x_{ \ell- 1}$ by the precubical
identity~\eqref{eq:pcub}, hence we can let $( x_1',\dots, x_m')$ be the
cube path with $x_j'= x_j$ for $j\ne \ell$ and $x_\ell'= \delta_{ k_2}^0
x_{ \ell+ 1}$.

If $k_2> k_3$, then similarly $\delta_{ k_3}^1 x_{ \ell- 1}= \delta_{
  k_3}^1 \delta_{ k_2}^0 x_\ell= \delta_{ k_2- 1}^0 \delta_{ k_3}^1
x_\ell= \delta_{ k_2- 1}^0 x_{ \ell+ 1}$, and we can let $x_j'= x_j$ for
$j\ne \ell$ and $x_\ell'= \delta_{ k_3}^1 x_{ \ell- 1}$.

For the remaining case $k_2= k_3$, we replace $x_{ \ell- 1}$ by another
cube of equal dimension first: If $k_1< k_2$, then $x_{ \ell- 2}=
\delta_{ k_1}^0 \delta_{ k_2}^0 x_\ell= \delta_{ k_2- 1}^0 \delta_{
  k_1}^0 x_\ell$, hence the cube path $( x_1'',\dots, x_m'')$ with
$x_j''= x_j$ for $j\ne \ell- 1$ and $x_{ \ell- 1}''= \delta_{ k_1}^0
x_\ell$ is adjacent to $( x_1,\dots, x_m)$, and $T( x_1'',\dots, x_m'')=
T( x_1,\dots, x_m)$.  For this new cube path, we have $x_{ \ell- 2}''=
\delta_{ k_2- 1}^0 x_{ \ell- 1}''$, $x_{ \ell- 1}''= \delta_{ k_1}^0
x_\ell''$, and $x_{ \ell+ 1}''= \delta_{ k_3}^1 x_\ell''$, and as $k_1<
k_3$, we can apply to the cube path $( x_1'',\dots, x_m'')$ the argument
for the case $k_2< k_3$ above.

If $k_1\ge k_2$, then $x_{ \ell- 2}= \delta_{ k_1}^0 \delta_{ k_2}^0
x_\ell= \delta_{ k_2}^0 \delta_{ k_1+ 1}^0 x_\ell$ by another
application of the precubical identity~\eqref{eq:pcub}.  Hence we can
let $x_j''=x_j$ for $j\ne \ell- 1$ and $x_{ \ell- 1}''= \delta_{ k_1+
  1}^0 x_\ell$.  Then $x_{ \ell- 2}''= \delta_{ k_2}^0 x_{ \ell- 1}''$,
$x_{ \ell- 1}''= \delta_{ k_1+ 1}^0 x_\ell''$, and $x_{ \ell+ 1}''=
\delta_{ k_3}^1 x_\ell''$, and as $k_1+ 1> k_3$, we can apply the
argument for the case $k_2> k_3$ above.  \qed

\proof[Proof of Theorem~\ref{th:univ_cov}]%
It is clear that the structure maps $\tilde \delta_k^1$ are
well-defined.  For showing that also the mappings $\tilde \delta_k^0$
are well-defined, we note first that $\tilde \delta_k^0[ x_1,\dots,x_m]$
is independent of the representative chosen for $[ x_1,\dots, x_m]$: If
$( x_1',\dots, x_m')\sim( x_1,\dots, x_m)$, then $( y_1,\dots, y_p)\in
\tilde \delta_k^0[ x_1',\dots, x_m']$ if and only if $y_p= \delta_k^0
x_m'= \delta_k^0 x_m$ and $( y_1,\dots, y_p, x_m')=( y_1,\dots, y_p,
x_m)\sim( x_1',\dots, x_m')\sim( x_1,\dots, x_m)$, if and only if $(
y_1,\dots, y_p)\in \tilde \delta_k^0[ x_1,\dots, x_m]$.

We are left with showing that $\tilde \delta_k^0[ x_1,\dots, x_m]$ is
non-empty.  By Lemma~\ref{le:fan} there is a fan-shaped cube path $(
x_1',\dots, x_m')\in[ x_1,\dots, x_m]$, and by
Lemma~\ref{le:hom_fullcube} we can assume that $x_{ m- 1}'= \delta_k^0
x_m'= \delta_k^0 x_m$, hence $( x_1',\dots, x_{ m- 1}')\in \tilde
\delta_k^0[ x_1,\dots, x_m]$.

We need to show the precubical identity $\tilde \delta_k^\nu \tilde
\delta_\ell^\mu= \tilde \delta_{ \ell- 1}^\mu \tilde \delta_k^\nu$ for
$k< \ell$ and $\nu, \mu\in\{ 0, 1\}$.  For $\nu= \mu= 1$ this is clear,
and for $\nu= \mu= 0$ one sees that $( y_1,\dots, y_p)\in \tilde
\delta_k^0 \tilde \delta_\ell^0[ x_1,\dots, x_m]$ if and only if $y_p=
\delta_k^0 \delta_\ell^0 x_m= \delta_{ \ell- 1}^0 \delta_k^0 x_m$ and $(
x_1,\dots, x_m)\sim ( y_1,\dots, y_p, \delta_\ell^0 x_m, x_m)\sim(
y_1,\dots, y_p, \delta_k^0 x_m, x_m)$, by adjacency.

The cases $\nu= 1$, $\mu= 0$ and $\nu= 0$, $\mu= 1$ are similar to each
other, so we only show the former.  Let $( x_1',\dots, x_m')\in[
x_1,\dots, x_m]$ be a fan-shaped cube path with $x_{ m- 1}'=
\delta_\ell^0 x_m'$, \cf~Lemma~\ref{le:hom_fullcube}.  Then $\tilde
\delta_k^1 \tilde \delta_\ell^0[ x_1,\dots, x_m]= \tilde \delta_k^1[
x_1',\dots, x_{ m- 1}']=[ x_1',\dots, x_{ m- 1}', \delta_k^1 x_{ m-
  1}']$.  Now $\delta_k^1 x_{ m- 1}'= \delta_k^1 \delta_\ell^0 x_m'=
\delta_{ \ell- 1}^0 \delta_k^1 x_m$, and by adjacency, $( x_1',\dots,
x_{ m- 1}', \delta_k^1 x_{ m- 1}', \delta_k^1 x_m')\sim( x_1',\dots, x_{
  m- 1}', x_m', \delta_k^1 x_m')$, so that we have $( x_1',\dots, x_{ m-
  1}', \delta_k^1 x_{ m- 1}')\in \tilde \delta_{ \ell- 1}^0[ x_1',\dots,
x_m', \delta_k^1 x_m']= \tilde \delta_{ \ell- 1}^0 \tilde \delta_k^1[
x_1',\dots, x_m']$.

For showing that the projection $\pi_X: \tilde X\to X$ is a precubical
morphism, we note first that $\pi_X \tilde \delta_k^1[ x_1,\dots, x_m]=
\pi_X[ x_1,\dots, x_m, \delta_k^1 x_m]= \delta_k^1 x_m= \delta_k^1
\pi_X[ x_1,\dots, x_m]$ as required.  For $\tilde \delta_k^0$, let again
$( x_1',\dots, x_m')\in[ x_1,\dots, x_m]$ be a fan-shaped cube path with
$x_{ m- 1}'= \delta_k^0 x_m'$.  Then $\pi_X \tilde \delta_k^0[
x_1,\dots, x_m]= \pi_X[ x_1',\dots, x_{ m- 1}']= x_{ m- 1}'= \delta_k^0
x_m'= \delta_k^0 x_m= \delta_k^0 \pi_X[ x_1,\dots, x_m]$.

The proof that $*\to \tilde X$ is a higher-dimensional tree follows from
Lemma~\ref{le:cube_path_rep} below: Let $( \tilde x_1,\dots, \tilde
x_m)$, $( \tilde y_1,\dots, \tilde y_m)$ be pointed cube paths in
$\tilde X$ with $\tilde x_m= \tilde y_m$, then we need to prove that $(
\tilde x_1,\dots, \tilde x_m)\sim( \tilde y_1,\dots, \tilde y_m)$.  Let
$x_j= \pi_X \tilde x_j$, $y_j= \pi_X \tilde y_j$ for $j= 1,\dots, m$ be
the projections, then $( x_1,\dots, x_m)$, $( y_1,\dots, y_m)$ are
pointed cube paths in $X$.  By Lemma~\ref{le:cube_path_rep}, $(
x_1,\dots, x_j)\in \tilde x_j$ and $( y_1,\dots, y_j)\in \tilde y_j$ for
all $j= 1,\dots, m$.

By $\tilde x_m= \tilde y_m$, we know that $( x_1,\dots, x_m)\sim(
y_1,\dots, y_m)$.  Let $( x_1,\dots, x_m)=( z^1_1,\dots,
z^1_m)\sim\cdots\sim( z^p_1,\dots, z^p_m)=( y_1,\dots, y_m)$ be a
sequence of adjacencies, and let $\tilde z^\ell_j=[ z^\ell_1,\dots,
z^\ell_j]$.  This defines pointed cube paths $( \tilde z^\ell_1,\dots,
\tilde z^\ell_m)$ in $\tilde X$; we show that $( \tilde x_1,\dots,
\tilde x_m)=( \tilde z^1_1,\dots, \tilde z^1_m)\sim\cdots\sim( \tilde
z^p_1,\dots, \tilde z^p_m)=( \tilde y_1,\dots, \tilde y_m)$ is a
sequence of adjacencies:

Let $\ell\in\{ 1,\dots,p- 1\}$, and let $\alpha\in\{ 1,\dots, m- 1\}$ be
the index such that $z^\ell_\alpha\ne z^{ \ell+ 1}_\alpha$ and
$z^\ell_j= z^{ \ell+ 1}_j$ for all $j\ne \alpha$.  Then $(
z^\ell_1,\dots, z^\ell_j)=( z^{ \ell+1}_1,\dots, z^{ \ell+ 1}_j)$ for
$j< \alpha$ and $( z^\ell_1,\dots, z^\ell_j)\sim( z^{ \ell+1}_1,\dots,
z^{ \ell+ 1}_j)$ for $j> \alpha$, hence there is an adjacency $( \tilde
z^\ell_1,\dots, \tilde z^\ell_m)\sim( \tilde z^{ \ell+ 1}_1,\dots,
\tilde z^{ \ell+ 1}_m)$.

For the last claim of the proposition, if $X$ itself is a
higher-dimensional tree, then an inverse to $\pi_X$ is given by mapping
$x\in X$ to the unique equivalence class $[ x_1,\dots, x_m]\in \tilde X$
of any pointed cube path $( x_1,\dots, x_m)$ in $X$ with $x_m= x$.  \qed

\begin{lem}
  \label{le:cube_path_rep}
  If $X$ is a higher-dimensional automaton and $( \tilde x_1,\dots,
  \tilde x_m)$ is a pointed cube path in $\tilde X$, then $( \pi_X
  \tilde x_1,\dots, \pi_X \tilde x_j)\in \tilde x_j$ for all $j=
  1,\dots, m$.
\end{lem}

\proof%
  Let $x_j= \pi_X \tilde x_j$, for $j= 1,\dots, m$, then $( x_1,\dots,
  x_m)$ is a pointed cube path in $X$.  We show the claim by induction:
  We have $\tilde x_1= \tilde i=[ i]=[ x_1]$, so assume that $(
  x_1,\dots, x_j)\in \tilde x_j$ for some $j\in\{ 1,\dots, m- 1\}$.  If
  $\tilde x_{ j+ 1}= \tilde \delta_k^1 \tilde x_j$ for some $k$, then
  $x_{ j+ 1}= \delta_k^1 x_j$, and $( x_1,\dots, x_{ j+ 1})\in \tilde
  x_{ j+ 1}$ by definition of $\tilde \delta_k^1$.  Similarly, if
  $\tilde x_j= \tilde \delta_k^0 \tilde x_{ j+ 1}$ for some $k$, then
  $x_j= \delta_k^0 x_{j+ 1}$, and $( x_1,\dots, x_{ j+ 1})\in \tilde x_{
    j+ 1}$ by definition of $\tilde \delta_k^0$.
\qed

\begin{lem}
  \label{le:unf_unique_lift}
  For any HDA $X$ there is a unique lift $r$ in any commutative diagram
  as below, for morphisms $g: P\to Q\in \HDP$, $p: P\to \tilde X, q:
  Q\to X\in \HDA$:
  \begin{equation*}
    \xymatrix{%
      P \ar[r]^p \ar[d]_g & \tilde X \ar[d]^{ \pi_X} \\ Q \ar[r]_q
      \ar@{.>}[ur]|r & X
    }
  \end{equation*}
\end{lem}

\proof%
  Let $( \tilde x_1,\dots, \tilde x_m)$ be a pointed cube path in
  $\tilde X$, and write $x_j= \pi_X \tilde x_j$ for $j= 1,\dots, m$.
  Let $( x_1,\dots,x_m, y_1,\dots, y_p)$ be an extension in $X$ and
  define $\tilde y_j=[ x_1,\dots, x_m, y_1,\dots, y_j]$ for $j= 1,\dots,
  p$.  Then $( \tilde x_1,\dots, \tilde x_m, \tilde y_1,\dots, \tilde
  y_p)$ is the required extension in $\tilde X$, which is unique as
  $\tilde X$ is a higher-dimensional tree.
\qed

\begin{cor}
  \label{cor:proj_open}
  Projections are open, and any HDA is hd-bisimilar to its
  unfolding. \qed
\end{cor}

\section{Higher-dimensional Automata up to Homotopy}
\label{se:hdah}

\begin{defi}
  \label{de:hdah}
  The category of \emph{higher-dimensional automata up to homotopy}
  $\HDAh$ has as objects HDA and as morphisms pointed precubical
  morphisms $f:\tilde X\to \tilde Y$ of unfoldings.
\end{defi}

\begin{figure}
  \centering
  \begin{tikzpicture}[->,>=stealth',auto,scale=.64]
    \tikzstyle{every node}=[font=\footnotesize,initial text=]
    \tikzstyle{every state}=[fill=white,shape=circle,inner
    sep=.5mm,minimum size=3mm]
    \begin{scope}
      \node[state,initial] (x) at (0,0) {};
      \node[state] (y) at (2,0) {};
      \path[out=-30,in=-150] (x) edge (y);
      \path[out=-210,in=30] (y) edge (x);
      \node at (1,-1) {$X$};
      \path (1,4.5) edge node {$\pi_X$} (1,1.5);
    \end{scope}
    \begin{scope}[yshift=5cm]
      \node[state,initial] (x0) at (0,0) {};
      \node[state] (y0) at (2,1) {};
      \node[state] (x1) at (0,2) {};
      \node[state] (y1) at (2,3) {};
      \node[state] (x2) at (0,4) {};
      \path (x0) edge (y0);
      \path (y0) edge (x1);
      \path (x1) edge (y1);
      \path (y1) edge (x2);
      \path[dotted] (x2) edge (2,5);
      \node at (1,5.5) {$\tilde X$};
      \path (3.5,2.5) edge node {$f$} (6.5,2.5);
    \end{scope}
    \begin{scope}[xshift=8cm]
      \node[state,initial] (x) at (0,0) {};
      \node[state] (y) at (4,0) {};
      \node[state] (z) at (2,1) {};
      \path (x) edge (y);
      \path (y) edge (z);
      \path (z) edge (x);
      \node at (2,-1) {$Y$};
      \path (2,4.5) edge node {$\pi_Y$} (2,1.5);
    \end{scope}
    \begin{scope}[xshift=8cm,yshift=5cm]
      \node[state,initial] (x0) at (0,0) {};
      \node[state] (y0) at (4,1) {};
      \node[state] (z0) at (2,2) {};
      \node[state] (x1) at (0,3) {};
      \node[state] (y1) at (4,4) {};
      \path (x0) edge (y0);
      \path (y0) edge (z0);
      \path (z0) edge (x1);
      \path (x1) edge (y1);
      \path[dotted] (y1) edge (2,5);
      \node at (2,5.5) {$\tilde Y$};
    \end{scope}
  \end{tikzpicture}
  \caption{%
    \label{fi:hda_vs_hdah}
    Two simple one-dimensional HDA as objects of $\HDA$ and $\HDAh$.  In
    $\HDA$ there is no morphism $X\to Y$, in $\HDAh$ there is precisely
    one morphism $f: X\to Y$.
  }
\end{figure}
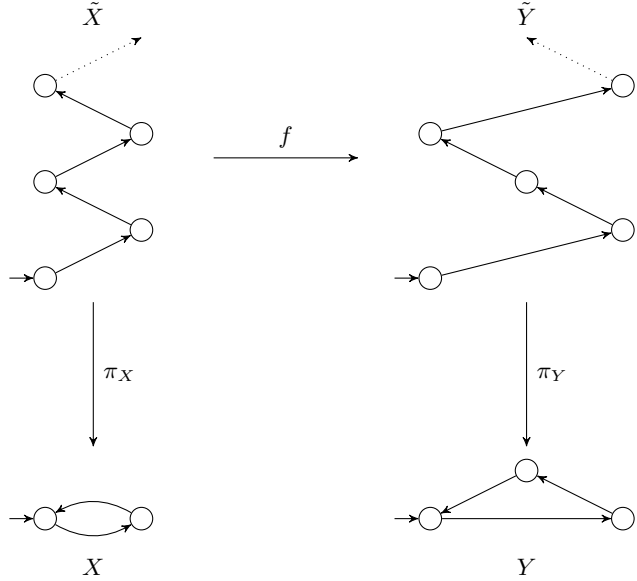

Hence any morphism $X\to Y$ in $\HDA$ gives, by the unfolding functor,
rise to a morphism $X\to Y$ in $\HDAh$.  The simple example in
Fig.~\ref{fi:hda_vs_hdah} shows that the converse is not the case.  By
restriction to higher-dimensional trees, we get a full subcategory
$\HDTh\hookrightarrow \HDAh$.

\begin{lem}
  \label{le:hdt_vs_hdth}
  The natural projection isomorphisms $\pi_X: \tilde X\to X$ for $X\in
  \HDT$ extend to an isomorphism of categories $\HDTh\cong \HDT$.
\end{lem}

\proof%
  Using the projection isomorphisms, any morphism $f: X\to Y$ in $\HDTh$
  can be ``pulled down'' to a morphism $\pi_Y\circ f\circ \pi_X^{ -1}:
  X\to Y$ of $\HDT$.
\qed

Restricting the above isomorphism to the subcategory $\HDP$ of $\HDT$
allows us to identify a subcategory $\HDPh$ of $\HDTh$ isomorphic to
$\HDP$.

Analogously to the coreflection between transition systems and
synchronization trees in~\cite{WinskelN95-Models}, we have a
coreflection between higher-dimensional automata and
higher-dimensional trees:

\begin{prop}
  \label{th:hda_vs_hdt}
  The functor $U: \HDA\to \HDT$ given on objects by mapping $X\in \HDA$
  to its unfolding $\tilde X$ and on morphisms by mapping $f: X\to Y$ to
  $\tilde f: \tilde X\to \tilde Y$ given by $\tilde f[ x_1,\dots, x_m]=[
  f( x_1),\dots, f( x_m)]$ is right adjoint to the forgetful functor
  $\HDT\hookrightarrow \HDA$.  The counit morphisms are the projections
  $\pi_X: \tilde X\to X$.
\end{prop}

\proof%
First, $U$ is indeed functorial, as $f$ maps adjacent cube paths $(
x_1,\dots, x_m)\sim( y_1,\dots, y_m)$ to cube paths $( f x_1,\dots, f
x_m)$, $( f y_1,\dots, f y_m)$ which are identical or adjacent, hence
$\tilde f: \tilde X\to \tilde Y$ is well-defined.

To show adjointness, we need to see that any pointed morphism $f: T\to
Y\in \pCub$ from a higher-dimensional tree $*\to T$ to a
higher-dimensional automaton $*\to Y$ factors uniquely as $f=
\pi_Y\circ g: T\to \tilde Y\to Y$.  This amounts to filling-in the
dotted arrow in the diagram
\begin{equation*}
  \xymatrix{%
    \tilde Y \ar[r]^{ \pi_Y} & Y \\ \tilde T \ar[u]^{ \tilde f}
    \ar[r]_{ \pi_T} & T\,. \ar[u]_f \ar@{.>}[ul]|g
  }
\end{equation*}
By Proposition~\ref{th:univ_cov}, $\pi_T$ has an inverse $\psi_T$,
hence $g= \tilde f\circ \psi_T$ is the unique filler. \qed

Note that by Proposition~\ref{th:univ_cov}, the unit morphisms are
isomorphisms, hence the above adjunction is indeed a coreflection.

The following is the analogue of Proposition~\ref{th:hda_vs_hdt} for
the homotopy categories, with a similar proof.  Note however that here,
$\Uh$ is an \emph{isomorphism} on morphisms.

\begin{prop}
  \label{th:hdah_vs_hdth}
  The forgetful functor $\HDTh\hookrightarrow \HDAh$ has a right adjoint
  $\Uh$ given on objects by mapping $X\in \HDAh$ to its unfolding
  $\tilde X$ and on morphisms by mapping $f: X\to Y$ to $\tilde f:
  \tilde X\to \tilde Y$.  The counit morphisms are the projections
  $\pi_X: \tilde X\to X$. \qed
\end{prop}

The unit morphisms are again isomorphisms, hence the adjunction is a
coreflection.

Combining the functors of Propositions~\ref{th:hda_vs_hdt} and
\ref{th:hdah_vs_hdth} with the isomorphism of
Lemma~\ref{le:hdt_vs_hdth}, we have the following diagram of categories
and coreflections.  Note that the adjunctions do not compose.
\begin{equation*}
  \xymatrix@C=4em{%
    {\HDAh} \ar@<.7ex>[r]^\Uh \ar@{}[r]|{\scriptscriptstyle \top} &
    {\HDTh} \ar@<.7ex>@{^(->}[l]^\Jh \ar@<.7ex>[r]
    \ar@{}[r]|{\scriptscriptstyle \cong} & {\HDT} \ar@<.7ex>[l]
    \ar@<.7ex>@{^(->}[r]^J \ar@{}[r]|{\scriptscriptstyle \bot} &
    {\HDA} \ar@<.7ex>[l]^U
  }
\end{equation*}

The endofunctor $J\circ U$ on $\HDA$, which maps objects and morphisms
to their unfoldings, splits into an adjunction between $\HDA$ and
$\HDAh$.  Its left part is the ``inclusion'' $\HDA\hookrightarrow
\HDAh$ which we already saw above.

\begin{prop}
  \label{pr:hdah_vs_hda}
  There is a coreflection $U_1: \HDAh\leftrightarrows \HDA: U_2$, with
  $U_1$ left and $U_2$ right adjoint given by $U_1( X)= \tilde X$ on
  objects, $U_1( f)= f$ on morphisms, $U_2( X)= X$ on objects, and $U_2(
  f)= \tilde f$ on morphisms.  The counit morphisms are the projections
  $\pi_X: \tilde X\to X$.
\end{prop}

\proof%
We need to see that any precubical morphism $f: \tilde X\to Y$ factors
uniquely as $f= \pi_Y\circ g: \tilde X\to \tilde Y\to Y$, but as $\tilde
X$ is a higher-dimensional tree, this is clear by the isomorphism $\pi_{
  \tilde X}: \ttilde X\to \tilde X$ in the diagram
\begin{equation*}
  \xymatrix{%
    \tilde Y \ar[r]^{ \pi_Y} & Y \\ \ttilde X \ar[u]_{ \tilde f}
    \ar[r]_{ \pi_{ \tilde X}} & \tilde X\,. \ar[u]^f \ar@{.>}[ul]|g
  }
\end{equation*}
\qed

\section{Homotopy Bisimilarity}
\label{se:open-hdah}

\begin{defi}
  A pointed morphism $f: X\to Y$ in $\HDAh$ is \emph{open} if it has
  the right lifting property with respect to $\HDPh$, \ie~if it is the
  case that there is a lift $r$ in any commutative diagram as below,
  for all morphism $g: P\to Q\in \HDPh$, $p: P\to X, q: Q\to Y\in
  \HDAh$:
  \begin{equation*}
    \xymatrix{%
      P \ar[r]^p \ar[d]_g & X \ar[d]^f \\ Q \ar[r]_q \ar@{.>}[ur]|r &
      Y
    }
  \end{equation*}
  HDA $X$, $Y$ are \emph{homotopy bisimilar} if there is $Z\in \HDAh$
  and a span of open maps $X\from Z\to Y$ in $\HDAh$.
\end{defi}

The connections between open maps in $\HDAh$ and open maps in $\HDA$ are
as follows.

\begin{lem}
  \label{le:openh=open}
  \label{le:liftopen}
  A morphism $f: X\to Y$ in $\HDAh$ is open if and only if $f: \tilde
  X\to \tilde Y$ is open as a morphism of $\HDA$.  If $g: X\to Y$ is
  open in $\HDA$, then so is $\tilde g: \tilde X\to \tilde Y$.
\end{lem}

\proof%
  For the forward implication of the first claim, let
  \begin{equation}
    \label{eq:small-diag}
    \vcenter{%
      \xymatrix{%
        P \ar[r]^p \ar[d]_g & \tilde X \ar[d]^f \\ Q \ar[r]_q & \tilde Y
      }}
  \end{equation}
  be a diagram in $\HDA$ with $g: P\to Q\in \HDP$; we need to find a
  lift $Q\to \tilde X$.

  Using the isomorphisms $\pi_P: \tilde P\to P$, $\pi_Q: \tilde Q\to Q$,
  we can extend this diagram to the left; note that $\tilde g: \tilde
  P\to \tilde Q$ is a morphism of $\HDP$:
  \begin{equation}
    \label{eq:wide_diag}
    \vcenter{%
      \xymatrix{%
        \tilde P \ar[r]_{ \cong} \ar[d]_{ \tilde g} \ar@/^3ex/[rr]^{ p'} & P
        \ar[r]_p \ar[d]_g & \tilde X \ar[d]^f \\ \tilde Q \ar[r]^{ \cong}
        \ar@/_3ex/[rr]_{ q'} & Q \ar[r]^q & \tilde Y
      }}
  \end{equation}

  Hence we have a diagram
  \begin{equation*}
    \xymatrix{%
      P \ar[r]^{ p'} \ar[d]_{ \tilde g} & X \ar[d]^f \\ Q \ar[r]_{ q'} &
      Y
    }
  \end{equation*}
  in $\HDAh$, and as $\tilde g: P\to Q$ is a morphism of $\HDPh$, we
  have a lift $r: Q\to X$ in $\HDAh$.  This gives a morphism $r: \tilde
  Q\to \tilde X\in \HDA$ in Diagram~\eqref{eq:wide_diag}, and by
  composition with the inverse of the isomorphism $\pi_Q: \tilde Q\to
  Q$, a lift $r': Q\to \tilde X\in \HDA$ in
  Diagram~\eqref{eq:small-diag}.

  For the back implication in the first claim, assume $f: \tilde X\to
  \tilde Y\in \HDA$ open and let
  \begin{equation*}
    \xymatrix{%
      P \ar[r]^p \ar[d]_g & X \ar[d]^f \\ Q \ar[r]_q & Y
    }
  \end{equation*}
  be a diagram in $\HDAh$ with $g: P\to Q\in \HDPh$; we need to find a
  lift $Q\to X$.  Transferring this diagram to the category $\HDA$, we
  have
  \begin{equation*}
    \xymatrix{%
      \tilde P \ar[r]^p \ar[d]_g & \tilde X \ar[d]^f \\ \tilde Q
      \ar[r]_q & \tilde Y
    }
  \end{equation*}
  and as $g: \tilde P\to \tilde Q$ is a morphism of $\HDP$, we get the
  required lift.

  To prove the second claim, let
  \begin{equation*}
    \xymatrix{%
      P \ar[r]^p \ar[d]_h & \tilde X \ar[d]^(.6){ \tilde g} \\ Q \ar[r]_q &
      \tilde Y
    }
  \end{equation*}
  be a diagram in $\HDA$ with $h: P\to Q\in \HDP$.  We can extend it
  using the projection morphisms:
  \begin{equation*}
    \xymatrix{%
      P \ar[r]^p \ar[d]_h & \tilde X \ar[r]^{ \pi_X} \ar[d]^(.6){ \tilde g} &
      X \ar[d]^g \\ Q \ar[r]_q & \tilde Y \ar[r]_{ \pi_Y} & Y
    }
  \end{equation*}
  Because $g$ is open in $\HDA$, we hence have a lift
  \begin{equation*}
    \xymatrix{%
      P \ar[r]^p \ar[d]_h & \tilde X \ar[r]^{ \pi_X} \ar[d]^(.6){ \tilde g} &
      X \ar[d]^g \\ Q \ar[r]_q \ar@{.>}[rru]_(.7)r & \tilde Y \ar[r]_{ \pi_Y}
      & Y
    }
  \end{equation*}
  and Lemma~\ref{le:unf_unique_lift} then gives the required lift $r'$
  in the diagram
  \begin{equation*}
    \xymatrix{%
      P \ar[r]^p \ar[d]_g & \tilde X \ar[d]^{ \pi_X} \\ Q \ar[r]_r
      \ar@{.>}[ur]^{r'} & X
    }
  \end{equation*}
\qed

\begin{exa}
  The morphism $f: X\to Y$ in Fig.~\ref{fi:hda_vs_hdah} is open in
  $\HDAh$, showing that $X$ and $Y$ are, as expected, homotopy
  bisimilar.
\end{exa}

We also need a lemma on prefixes in unfoldings.

\begin{lem}
  \label{le:reach-vs-prefix}
  Let $X$ be a HDA and $\tilde x, \tilde z\in \tilde X$.  Then there is
  a cube path from $\tilde x$ to $\tilde z$ in $\tilde X$ if and only if
  $\tilde x\sqsubseteq \tilde z$.
\end{lem}

\proof%
  For the forward implication, let $( \tilde x, \tilde y_1,\dots, \tilde
  y_p)$ be a cube path in $\tilde X$ with $\tilde y_p= \tilde z$, let $(
  x_1,\dots, x_m)\in \tilde x$, and write $y_j= \pi_X \tilde y_j$ for
  all $j$.  By Lemma~\ref{le:cube_path_rep}, $( x_1,\dots, x_m,
  y_1,\dots, y_p)\in \tilde z$.

  For the other direction, let $( x_1,\dots, x_m, y_1,\dots, y_p)\in
  \tilde z$ such that $( x_1,\dots, x_m)\in \tilde x$, and define
  $\tilde y_j=[ x_1,\dots, x_m, y_1,\dots, y_j]$ for all $j$.  Then $(
  \tilde x, \tilde y_1,\dots, \tilde y_p)$ is the required cube path
  from $\tilde x$ to $\tilde z$ in $\tilde X$.
\qed

\begin{prop}
  \label{th:hombis}
  For HDA $i: *\to X$, $j: * \to Y$, the following are equivalent:
  \begin{enumerate}
  \item\label{enu:hombis.box} $X$ and $Y$ are homotopy bisimilar;
  \item\label{enu:hombis.onestep} there exists a precubical subset
    $\tilde R\subseteq \tilde X\times \tilde Y$ with $( \tilde i, \tilde
    j)\in \tilde R$, and such that for all $( \tilde x_1, \tilde y_1)\in
    \tilde R$,
    \begin{itemize}
    \item for any $\tilde x_2\in \tilde X$ for which $\tilde x_1=
      \delta_k^0 \tilde x_2$ for some $k$, there exists $\tilde y_2\in
      \tilde Y$ for which $\tilde y_1= \delta_k^0 \tilde y_2$ and $(
      \tilde x_2, \tilde y_2)\in \tilde R$,
    \item for any $\tilde y_2\in \tilde Y$ for which $\tilde y_1=
      \delta_k^0 \tilde y_2$ for some $k$, there exists $\tilde x_2\in
      \tilde X$ for which $\tilde x_1= \delta_k^0 \tilde x_2$ and $(
      \tilde x_2, \tilde y_2)\in \tilde R$;
    \end{itemize}
  \item\label{enu:hombis.cubepath} there exists a precubical subset
    $\tilde R\subseteq \tilde X\times \tilde Y$ with $( \tilde i, \tilde
    j)\in \tilde R$, and such that for all $( \tilde x_1, \tilde y_1)\in
    \tilde R$,
    \begin{itemize}
    \item for any cube path $( \tilde x_1,\dots, \tilde x_n)$ in $\tilde
      X$, there exists a cube path $( \tilde y_1,\dots, \tilde y_n)$ in
      $\tilde Y$ with $( \tilde x_p, \tilde y_p)\in \tilde R$ for all
      $p= 1,\dots, n$,
    \item for any cube path $( \tilde y_1,\dots, \tilde y_n)$ in $\tilde
      Y$, there exists a cube path $( \tilde x_1,\dots, \tilde x_n)$ in
      $\tilde X$ with $( \tilde x_p, \tilde y_p)\in \tilde R$ for all
      $p= 1,\dots, n$;
    \end{itemize}
  \item\label{enu:hombis.prefix} there exists a precubical subset
    $\tilde R\subseteq \tilde X\times \tilde Y$ with $( \tilde i, \tilde
    j)\in \tilde R$, and such that for all $( \tilde x_1, \tilde y_1)\in
    \tilde R$,
    \begin{itemize}
    \item for any $\tilde x_2\sqsupseteq \tilde x_1$ in $\tilde X$,
      there exists $\tilde y_2\sqsupseteq \tilde y_1$ in $\tilde Y$ for
      which $( \tilde x_2, \tilde y_2)\in \tilde R$,
    \item for any $\tilde y_2\sqsupseteq \tilde y_1$ in $\tilde Y$,
      there exists $\tilde x_2\sqsupseteq \tilde x_1$ in $\tilde X$ for
      which $( \tilde x_2, \tilde y_2)\in \tilde R$.
    \end{itemize}
  \end{enumerate}
\end{prop}

Again, the requirement that $\tilde R$ be a precubical subset is
equivalent to saying that whenever $( \tilde x, \tilde y)\in \tilde R$,
then also $( \delta_k^\nu \tilde x, \delta_k^\nu \tilde y)\in \tilde R$
for any $k$ and $\nu\in\{ 0, 1\}$.

\proof%
  The implication
  \eqref{enu:hombis.box}~$\Longrightarrow$~\eqref{enu:hombis.onestep}
  follows directly from Theorem~\ref{th:bisim},
  and~\eqref{enu:hombis.cubepath} can be proven
  from~\eqref{enu:hombis.onestep} by induction.
  % (We can omit the reachability condition from
  % items~\eqref{enu:hombis.onestep} and~\eqref{enu:hombis.cubepath}
  % because any cube in an unfolding is reachable.)
  Equivalence of~\eqref{enu:hombis.cubepath}
  and~\eqref{enu:hombis.prefix} is immediate from
  Lemma~\ref{le:reach-vs-prefix}.

  For the implication
  \eqref{enu:hombis.cubepath}~$\Longrightarrow$~\eqref{enu:hombis.box},
  we can use Theorem~\ref{th:bisim} to get a span $\tilde X\tfrom f
  R\tto g \tilde Y$ of open maps in $\HDA$.  Connecting these with the
  projection $\pi_R: \tilde R\to R$ gives a span $\tilde X\tfrom{ f\circ
    \pi_R} \tilde R\tto{ g\circ \pi_R} \tilde Y$.  By
  % Corollaries~\ref{cor:open*open=open} and
  Corollary~\ref{cor:proj_open}, the maps in the span are open in
  $\HDA$, hence by Lemma~\ref{le:openh=open}, $X\tfrom{ f\circ \pi_R}
  R\tto{ g\circ \pi_R} Y$ is a span of open maps in $\HDAh$.
\qed

\begin{thm}
  \label{th:hbis=bis}
  HDA $X$, $Y$ are homotopy bisimilar if and only if they are
  hd-bisimilar.
\end{thm}

\proof%
A span of open maps $X\tfrom f Z\tto g Y$ in $\HDA$ lifts to a span
$X\tfrom{ \tilde f} Z\tto{ \tilde g} Y$ in $\HDAh$, and $\tilde f$ and
$\tilde g$ are open by Lemma~\ref{le:liftopen}.  Hence hd-bisimilarity
implies homotopy bisimilarity.

For the other direction, let $X\tfrom f Z\tto g Y$ be a span of open
maps in $\HDAh$.  In $\HDA$, this is a span $\tilde X\tfrom f \tilde
Z\tto g \tilde Y$, and composing with the projections yields $X\tfrom{
  \pi_X\circ f} \tilde Z\tto{ \pi_Y\circ g} Y$.  By
Lemma~\ref{le:openh=open} and Corollary~\ref{cor:proj_open}, both
$\pi_x\circ f$ and $\pi_Y\circ g$ are open in $\HDA$.  \qed

\begin{cor}
  \label{co:decidable}
  Homotopy bisimilarity is decidable for finite HDA.
\end{cor}

\proof%
  The condition in Thm.~\ref{th:bisim}\eqref{enu:bisim.onestep}
  immediately gives rise to a fixed-point algorithm similar to the one
  used to decide standard bisimilarity, \cf~\cite{book/Milner89}.
\qed

In order to be able to relate our notion of bisimilarity to other common
notions in Section~\ref{se:comparison} below, we translate it to a
relation between pointed cube paths, \ie~executions:

\begin{thm}
  \label{th:biscp}
  HDA $i: *\to X$, $j: *\to Y$ are homotopy bisimilar if and only if
  there exists a relation $R$ between pointed cube paths in $X$ and
  pointed cube paths in $Y$ for which $(( i),( j))\in R$, and such that
  for all $( \rho, \sigma)\in R$ with $\rho=( x_1,\dotsc, x_m)$ and
  $\sigma=( y_1,\dotsc, y_p)$,
  \begin{itemize}
  \item $\dim x_m= \dim y_p$,
  \item for all $k= 1,\dots, \dim x_m$, $( \rho* \delta_k^1 x_m, \sigma*
    \delta_k^1 y_p)\in R$,
  \item for all $k= 1,\dots, \dim x_m$, there exist $\rho'\in \tilde
    \delta_k^0[ \rho]$ and $\sigma'\in \tilde \delta_k^0[ \sigma]$ with
    $( \rho', \sigma')\in R$,
  \item for all $\rho'\sim \rho$, there exists $\sigma'\sim \sigma$
    with $( \rho', \sigma')\in R$,
  \item for all $\sigma'\sim \sigma$, there exists $\rho'\sim \rho$
    with $( \rho', \sigma')\in R$,
  \item for all $\rho'\sqsupseteq \rho$, there exists
    $\sigma'\sqsupseteq \sigma$ with $( \rho', \sigma')\in R$,
  \item for all $\sigma'\sqsupseteq \sigma$, there exists
    $\rho'\sqsupseteq \rho$ with $( \rho', \sigma')\in R$.
  \end{itemize}
\end{thm}

Note how the last four conditions are reminiscent of the ones for
\emph{history-preserving
  bisimilarity}~\cite{DBLP:journals/tcs/Glabbeek06}.

\proof%
For the ``if'' part of the theorem, assume that we have a relation $R$
as in the theorem and define $\tilde R\subseteq \tilde X\times \tilde Y$
by $\tilde R=\{([ \rho],[ \sigma])\mid( \rho, \sigma)\in R\}$.  Then $(
\tilde i, \tilde j)\in \tilde R$, and the first three conditions ensure
that $\tilde R$ is a precubical subset of $\tilde X\times \tilde Y$: By
$\dim x_m= \dim y_p$, $\tilde R_n\subseteq \tilde X_n\times \tilde Y_n$
for all $n$, the second condition implies that for all $([ \rho],[
\sigma])\in \tilde R$ and all $k$, also $\tilde \delta_k^1([ \rho],[
\sigma])=([ \rho* \delta_k^1 x_m],[ \sigma* \delta_k^1 y_m])\in \tilde
R$, and using the third condition, also $\tilde \delta_k^0([ \rho],[
\sigma])=([ \rho'],[ \sigma'])\in \tilde R$.

Now let $( \tilde x_1, \tilde y_1)\in \tilde R$ and $\tilde
x_2\sqsupseteq \tilde x_1$.  We have $\rho_1\in \tilde x_1$ and
$\sigma_1\in \tilde y_1$ for which $( \rho_1, \sigma_1)\in R$.  Let
$\rho_1'\in \tilde x_1$ and $\rho_2\in \tilde x_2$ such that
$\rho_2\sqsupseteq \rho_1'$, then $\rho_1'\sim \rho_1$, hence we have
$\sigma_1'\sim \sigma_1$ for which $( \rho_1', \sigma_1')\in R$.  By
$\rho_2\sqsupseteq \rho_1'$ we also have $\sigma_2\sqsupseteq \sigma_1'$
for which $( \rho_2, \sigma_2)\in R$, hence $( \tilde x_2=[ \rho_2],[
\sigma_2])\in \tilde R$ as was to be shown.  The symmetric condition in
Theorem~\ref{th:hombis}\eqref{enu:hombis.prefix} can be shown
analogously.

For the other implication, let $\tilde R\subseteq \tilde X\times \tilde
Y$ be a precubical subset as in
Theorem~\ref{th:hombis}\eqref{enu:hombis.prefix} and define a relation
of pointed cube paths by $R=\{( \rho, \sigma)\mid([ \rho],[ \sigma])\in
\tilde R\}$.  Then $(( i),( j))\in R$.  Let $( \rho=( x_1,\dotsc, x_m),
\sigma=( y_1,\dotsc, y_p))\in R$, then $\dim x_m= \dim y_p$ by $\tilde
R_n\subseteq \tilde X_n\times \tilde Y_n$.  Let $k\in\{ 1,\dotsc, \dim
x_m\}$, then $\tilde \delta_k^1([ \rho],[ \sigma])\in \tilde R$ and
hence $( \rho* \delta_k^1 x_m, \sigma* \delta_k^1 y_p)\in R$.  Using
$\tilde \delta_k^0([ \rho],[ \sigma])\in \tilde R$, we see that there
must exist $\rho'\in \tilde \delta_k^0[ \rho]$ and $\sigma'\in \tilde
\delta_k^0[ \sigma]$ with $( \rho', \sigma')\in R$.

Now let $( \rho, \sigma)\in R$, then also $( \rho', \sigma')\in R$ for
any $\rho'\sim \rho$, $\sigma'\sim \sigma$, showing the fourth and fifth
conditions of the theorem.  For the sixth one, let $\rho'\sqsupseteq
\rho$, then $[ \rho']\sqsupseteq[ \rho]$, hence we have $\tilde
y_2\sqsupseteq[ \sigma]$ for which $([ \rho'], \tilde y_2)\in \tilde R$.
By definition of $R$ we have $( \rho', \sigma')\in R$ for any
$\sigma'\in \tilde y_2$, and by $\tilde y_2\sqsupseteq[ \sigma]$, there
is $\sigma'\in \tilde y_2$ for which $\sigma'\sqsupseteq \sigma$,
showing the sixth condition.  The seventh condition is proved
analogously.  \qed

\section{Labels}
\label{se:labels}

% For labeling HDA, we need a subcategory of $\pCub$ isomorphic to the
% category of sets and functions.  For a (finite or infinite) set $S$, we
% construct a precubical set $\bang S=\{ \bang S_n\}$ by letting $\bang
% S_n= S^n$, with face maps defined by 
% % $\delta_k^\nu( a_1,\dotsc, a_n)=( a_1,\dotsc, a_{ n- k}, a_{ n- k+
% % 2},\dotsc, a_n)$.
% $\delta_k^\nu( a_1,\dotsc, a_n)=( a_1,\dotsc, a_{k- 1}, a_{ k+
%   1},\dotsc, a_n)$.

For labeling HDA, we need a subcategory of $\pCub$ isomorphic to the
category of sets and functions.  Given a finite or countably infinite
set $S=\{ a_1, a_2,\dots\}$, we construct a precubical set $\bang S=\{
\bang S_n\}$ by letting
\begin{equation*}
  \bang S_n=\big\{( a_{ i_1},\dots, a_{ i_n})\mid
  i_k\le i_{ k+ 1}\text{ for all } k= 1,\dots,n- 1\big\}
\end{equation*}
with face maps defined by $\delta_k^\nu( a_{ i_1},\dots, a_{ i_n})=(
a_{ i_1},\dots, a_{ i_{ k- 1}}, a_{ i_{ k+ 1}},\dots, a_{ i_n})$.

\begin{defi}
  The category of \emph{higher-dimensional tori} $\HDAt$ is the full
  subcategory of $\pCub$ generated by the objects $\bang S$.
\end{defi}

As any object in $\HDAt$ has precisely one $0$-cube, the pointed
category $*\downarrow \HDAt$ is isomorphic to $\HDAt$.  Note that the
objects in $\HDAt$ indeed are tori: by definition, lower and upper
boundaries of any $n$-cube agree, hence all $n$-cubes are loops.

\begin{lem}
  $\HDAt$ is isomorphic to the category of sets and functions.
\end{lem}

\proof%
A function $f: S\to T$ is lifted to $\bang f: \bang S\to \bang T$ by $f(
a_1,\dotsc, a_n)= \langle f( a_1),\dotsc, f( a_n)\rangle$, where the
elements on the right-hand side are re-sorted.  This is easily seen to
be a precubical mapping.  The inverse direction follows from the fact
that the objects in $\HDAt$ are \emph{coskeletal} on their $1$-cubes,
\cf~\cite{Fahrenberg05-hda, bh81}.  \qed

\begin{defi}
  The category of \emph{labeled higher-dimensional automata} is the
  pointed arrow category $\LHDA= *\downarrow \pCub\to \HDAt$, with
  objects $*\to X\to \bang S$ labeled pointed precubical sets and
  morphisms commutative diagrams
  \begin{equation*}
    \xymatrix@C=1.5em@R=1.2em{%
      & {*} \ar[dl] \ar[dr] \\ X \ar[rr]_f \ar[d] && Y \ar[d] \\ {\bang
        S} \ar[rr]_\sigma && {\bang T}
    }
  \end{equation*}
\end{defi}

\begin{rem}
  If morphisms of labeled higher-dimensional automata are to model
  (functional) \emph{simulations}, then one needs \emph{partial}
  labeling morphisms $\sigma$.  This can be achieved by introducing
  \emph{degeneracies} for precubical sets, passing to the category
  $\Cub$ of \emph{cubical} sets.  One can then show that the full
  subcategory of $\Cub$ spanned by free cubical sets on
  higher-dimensional tori is isomorphic to the category of finite sets
  and partial functions and define $\LHDA$ accordingly.  This is indeed
  the approach taken in~\cite{Goubault02-cmcim,Fahrenberg05-hda}.  As we
  are only concerned with bisimilarity here, we do not need partial
  labeling morphisms.
\end{rem}

We now fix a labeling set $\Sigma$; we will work in the category with
morphisms
\begin{equation*}
  \xymatrix@C=1.5em@R=1.2em{%
    & {*} \ar[dl] \ar[dr] \\ X \ar[rr]_f \ar[d] && Y \ar[d] \\ {\bang
      \Sigma} \ar[rr]_\id && {\bang \Sigma}
  }
\end{equation*}

\begin{defi}
  A morphism $( f, \id):( *\to X\to \bang \Sigma)\to( *\to Y\to \bang
  \Sigma)$ in $\LHDA$ is \emph{open} if its component $f$ is open in
  $\HDA$.  Labeled HDA $*\to X\to \bang \Sigma$, $*\to Y\to \bang
  \Sigma$ are \emph{hd-bisimilar} if there is $*\to Z\to \bang \Sigma\in
  \LHDA$ and a span of open maps $X\from Z\to Y$ in $\LHDA$.
\end{defi}

% By definition, the torus $\bang \Sigma$ contains all $n$-cubes $(
% a_1,\dotsc, a_n)$.  Hence we have the following lemma:

% \begin{lem}
%   \label{le:cube_path_hom_bang}
%   Let $( x_1,\dots, x_m)$, $( y_1,\dots, y_m)$ be pointed cube paths in
%   $\bang \Sigma$ with $x_m= y_m$.  Then $( x_1,\dots, x_m)\sim ( y_1,\dots,
%   y_m)$.\uli{Sure?} \qed
% \end{lem}

% \begin{cor}
%   The unfolding of a higher-dimensional torus $i: *\to \bang \Sigma\in
%   \HDAt$ is isomorphic to the pointed precubical set $j: *\to Y$ given
%   as follows:\uli{Sure?}
%   \begin{itemize}
%   \item $Y_n=\{( x, m)\mid x\in \bang \Sigma_n, m\ge n, m\equiv n\mod 2\}$,
%     $j=( i, 0)$
%   \item $\delta_k^0( x, m)=( \delta_k^0 x, m- 1)$, $\delta_k^1( x, m)=(
%     \delta_k^1 x, m+ 1)$ \qed
%   \end{itemize}
% \end{cor}

The definitions of open maps and bisimilarity in $\HDAh$ can now easily
be extended to the labeled case.  Again, we will only need
label-preserving morphisms.

\begin{defi}
  The category of \emph{labeled higher-dimensional automata up to
    homotopy} $\LHDAh$ has as objects labeled HDA $*\to X\to \bang S$
  and as morphisms pairs of precubical morphisms $( f, \sigma):( *\to
  \tilde X\to \bang \tilde S)\to( *\to \tilde Y\to \bang \tilde T)$ of
  unfoldings.
\end{defi}

\begin{defi}
  A morphism $( f, \id):( *\to X\to \bang \Sigma)\to( *\to Y\to \bang
  \Sigma)$ in $\LHDAh$ is \emph{open} if its component $f$ is open in
  $\HDAh$.  Labeled HDA $*\to X\to \bang \Sigma$, $*\to Y\to \bang
  \Sigma$ are \emph{homotopy bisimilar} if there is a labeled HDA $*\to
  Z\to \bang \Sigma$ and a span of open maps $X\from Z\to Y$ in
  $\LHDAh$.
\end{defi}

As a corollary, we see that $*\to X\tto \lambda \bang \Sigma$, $*\to
Y\tto \mu \bang \Sigma$ are homotopy bisimilar if and only if there
exists a precubical subset $\tilde R\subseteq \tilde X\times \tilde Y$
like in Theorem~\ref{th:hombis} which \emph{respects homotopy classes of
  labels}, \ie~for which $\tilde \lambda( \tilde x)= \tilde \mu( \tilde
y)$ for each $( \tilde x, \tilde y)\in \tilde R$.

The proof of the next theorem is exactly the same as the one for
Theorem~\ref{th:hbis=bis}.

\begin{thm}
  Labeled HDA $X$, $Y$ are homotopy bisimilar if and only if they are
  hd-bisimilar.  \qed
\end{thm}

\section{Relation to Other Equivalences}
\label{se:comparison}

It remains to be seen how our homotopy bisimilarity relates to other
notions of equivalence for concurrent systems.

For a labeled HDA $*\to X\tto \lambda \bang \Sigma$, we extend $\lambda$
% and $\cptost$
to cube paths in $X$ by $\lambda( x_1,\dotsc, x_m)=(
\lambda( x_1),\dotsc, \lambda( x_m))$.
% and $\cptost( \rho)= \cptost( \lambda( \rho))$.

The following is a labeled version of Theorem~\ref{th:biscp}.

\begin{thm}
  \label{th:biscp-l}
  Labeled HDA $*\tto i X\tto \lambda \bang \Sigma$, $*\tto j Y\tto \mu
  \bang \Sigma$ are homotopy bisimilar if and only if there exists a
  relation $R$ between pointed cube paths in $X$ and pointed cube paths
  in $Y$ for which $(( i),( j))\in R$, and such that for all $( \rho,
  \sigma)\in R$ with $\rho=( x_1,\dotsc, x_m)$ and $\sigma=( y_1,\dotsc,
  y_p)$,
  \begin{itemize}
  \item for all $k= 1,\dots, \dim x_m$, $( \rho* \delta_k^1 x_m, \sigma*
    \delta_k^1 y_p)\in R$,
  \item for all $k= 1,\dots, \dim x_m$, there exist $\rho'\in \tilde
    \delta_k^0[ \rho]$ and $\sigma'\in \tilde \delta_k^0[ \sigma]$ with
    $( \rho', \sigma')\in R$,
  \item $\lambda( \rho)\sim \mu( \sigma)$,
  \item for all $\rho'\sim \rho$, there exists $\sigma'\sim \sigma$
    with $( \rho', \sigma')\in R$,
  \item for all $\sigma'\sim \sigma$, there exists $\rho'\sim \rho$
    with $( \rho', \sigma')\in R$,
  \item for all $\rho'\sqsupseteq \rho$, there exists
    $\sigma'\sqsupseteq \sigma$ with $( \rho', \sigma')\in R$,
  \item for all $\sigma'\sqsupseteq \sigma$, there exists
    $\rho'\sqsupseteq \rho$ with $( \rho', \sigma')\in R$.
  \end{itemize}
\end{thm}

\proof%
For the ``if'' part of the theorem, assume that we have a relation $R$
as in the theorem and define $\tilde R\subseteq \tilde X\times \tilde Y$
by $\tilde R=\{([ \rho],[ \sigma])\mid( \rho, \sigma)\in R\}$, as in the
proof of Theorem~\ref{th:biscp}.  Let $( \rho, \sigma)\in R$ and write
$\rho=( x_1,\dotsc, x_m)$ and $\sigma=( y_1,\dotsc, y_p)$.  By $\lambda(
\rho)\sim \mu( \sigma)$, also $\lambda( x_m)= \mu( y_p)$, which, as
$\lambda$ and $\mu$ are precubical mappings, implies that $\dim x_m=
\dim y_p$.

Thus $R$ satisfies the conditions of Theorem~\ref{th:biscp}, so we can
infer that $\tilde R\subseteq \tilde X\times \tilde Y$ is a precubical
subset for which the conditions in
Theorem~\ref{th:hombis}\eqref{enu:hombis.prefix} hold.  Let $([ \rho],[
\sigma])\in \tilde R$, then $\lambda( \rho)\sim \mu( \sigma)$ entails
$\tilde \lambda[ \rho]= \tilde \mu[ \sigma]$.

For the other direction, let $\tilde R\subseteq \tilde X\times \tilde Y$
be a precubical subset as in
Theorem~\ref{th:hombis}\eqref{enu:hombis.prefix} which respects labels.
Define a relation of pointed cube paths by $R=\{( \rho, \sigma)\mid([
\rho],[ \sigma])\in \tilde R\}$, then $R$ satisfies the conditions of
Theorem~\ref{th:biscp}.  Let $( \rho, \sigma)\in R$, then $([ \rho],[
\sigma])\in \tilde R$ implies $\tilde \lambda[ \rho]= \tilde \mu[
\sigma]$, hence $\lambda( \rho)\sim \mu( \sigma)$. \qed

\begin{thm}
  Homotopy bisimilarity is not implied by ST-bisimilarity and 
  incomparable with history-preserving bisimilarity.
\end{thm}

\proof%
This will follow from the examples below.  \qed

We finish this section by exposing several examples.  The first two
serve to position homotopy bisimilarity with regard to
history-preserving bisimilarity, and the last shows a case in which
homotopy bisimilarity distinguishes auto-concurrency in a way similar to
ST-bisimilarity.  Whether homotopy bisimilarity implies ST-bisimilarity,
and whether it is implied by hereditary history-preserving (hhp)
bisimilarity, is open.

\begin{figure}
  \centering
  \begin{tikzpicture}[auto,scale=1.2,->]
    \tikzstyle{every node}=[font=\footnotesize,initial text=]
    \tikzstyle{every state}=[fill=white,shape=circle,inner
    sep=.3mm,minimum size=2mm]
    \begin{scope}
      \path[fill=black!15] (0,0) to (-1,-1) to (0,-2) to (1,-1);
      \node[state,initial above] (x0) at (0,0) {$x_0$};
      \node[state] (x1) at (-1,-1) {$x_1$};
      \node[state] (x2) at (1,-1) {$x_2$};
      \node[state] (x3) at (-2,-2) {$x_3$};
      \node[state] (x4) at (0,-2) {$x_4$};
      \path (x0) edge node[right,pos=.6] {$y_1$} node[left]
      {\textcolor{red}{$a$}} (x1);
      \path (x0) edge node[left,pos=.6] {$y_2$} node[right]
      {\textcolor{red}{$\,b$}} (x2);
      \path (x1) edge node[right] {$y_3$} node[left,pos=.4]
      {\textcolor{red}{$b\,$}} (x3);
      \path (x1) edge node[left] {$y_4$} (x4);
      \path (x2) edge node[right] {$y_5$} (x4);
      \node (z) at (0,-1) {$z$};
    \end{scope}
    \begin{scope}[xshift=12em]
      \path[fill=black!15] (0,0) to (-1,-1) to (0,-2) to (1,-1);
      \node[state,initial above] (x0) at (0,0) {$x_0'$};
      \node[state] (x1) at (-1,-1) {$x_1'$};
      \node[state] (x2) at (1,-1) {$x_2'$};
      \node[state] (x4) at (0,-2) {$x_4'$};
      \path (x0) edge node[right,pos=.6] {$y_1'$} node[left]
      {\textcolor{red}{$a$}} (x1);
      \path (x0) edge node[left,pos=.6] {$y_2'$} node[right]
      {\textcolor{red}{$\,b$}} (x2);
      \path (x1) edge node[left] {$y_4'$} (x4);
      \path (x2) edge node[right] {$y_5'$} (x4);
      \node (z) at (0,-1) {$z'$};
    \end{scope}
  \end{tikzpicture}
  \caption{%
    \label{fi:hdbex1} Two HDA pertaining to Example~\ref{ex:hdbex1}.
    % For the HDA on the
    % left, $\lambda( y_1)= \lambda( y_5)= a$ and $\lambda( y_2)=
    % \lambda(
    % y_3)= \lambda( y_4)= b$; on the right, $\lambda'( y_1')= \lambda'(
    % y_5')= a$ and $\lambda'( y_2')= \lambda'( y_4')= b$.
  }
\end{figure}
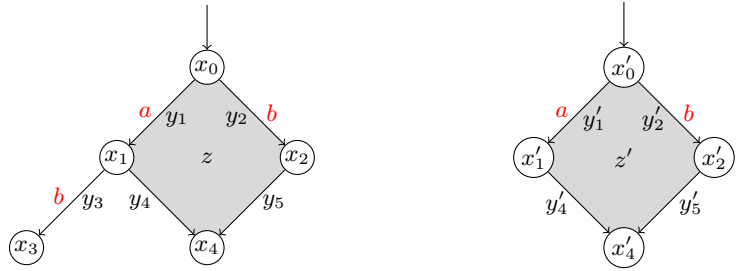

\begin{exa}
  \label{ex:hdbex1}
  The two HDA in Fig.~\ref{fi:hdbex1} are hd-bisimilar, as witnessed by
  the following precubical subset $R\subseteq X\times X'$:
  \begin{align*}
    R_0 &= \{( x_0, x_0'),( x_1, x_1'),( x_2, x_2'),( x_3, x_4'),( x_4,
    x_4')\} \\
    R_1 &= \{( y_1, y_1'),( y_2, y_2'),( y_3, y_4'),( y_4, y_4'),( y_5,
    y_5')\} \\
    R_2 &= \{( z, z')\}
  \end{align*}
  In~\cite[Example 5.2.2]{DBLP:conf/parle/GlabbeekV87} it is shown that
  the Petri-net translations of these HDA are ST-bisimilar, but not
  history-preserving bisimilar.
\end{exa}

\begin{figure}
  \centering
  \begin{tikzpicture}[auto,scale=1.4,->]
    \tikzstyle{every node}=[font=\footnotesize,initial text=]
    \tikzstyle{every state}=[fill=white,shape=circle,inner
    sep=.3mm,minimum size=6mm]
    \begin{scope}
      \path[fill=black!15] (0,0) to (-1,1) to (-3,1) to (-2,0);
      \path[fill=black!15] (0,0) to (1,1) to (3,1) to (2,0);
      \path[fill=black!15] (0,0) to (-1,-1) to (-3,-1) to (-2,0);
      \path[fill=black!15] (0,0) to (1,-1) to (3,-1) to (2,0);
      \path[fill=black!15] (0,0) to (-.8,-2) to (0,-3) to (.8,-2);
      \node[state,initial above] (x0) at (0,0) {$x_0$};
      \node[state] (x1) at (-1,1) {$x_1$};
      \node[state] (x2) at (-2,0) {$x_2$};
      \node[state] (x3) at (-1,-1) {$x_3$};
      \node[state] (x4) at (-3,1) {$x_4$};
      \node[state] (x5) at (-3,-1) {$x_5$};
      \path (x0) edge node[left] {$y_1$} (x1);
      \path (x0) edge node[above] {$y_2$} (x2);
      \path (x0) edge node[left] {$y_3$} (x3);
      \path (x1) edge node[below] {$y_4$} node[above]
      {\textcolor{red}{$b$}} (x4);
      \path (x2) edge node[right] {$y_5$} node[left]
      {\textcolor{red}{$c\,$}} (x4);
      \path (x2) edge node[right] {$y_6$} node[left]
      {\textcolor{red}{$a\,$}} (x5);
      \path (x3) edge node[below] {$y_7$} (x5);
      \node (z1) at (-1.5,.5) {$z_1$};
      \node (z2) at (-1.5,-.5) {$z_2$};
      \node[state] (x6) at (1,1) {$x_6$};
      \node[state] (x7) at (2,0) {$x_7$};
      \node[state] (x8) at (1,-1) {$x_8$};
      \node[state] (x9) at (3,1) {$x_9$};
      \node[state] (x10) at (3,-1) {$x_{10}$};
      \path (x0) edge node[right] {$y_8$} (x6);
      \path (x0) edge node[above] {$y_9$} (x7);
      \path (x0) edge node[right] {$y_{10}$} (x8);
      \path (x6) edge node[below] {$y_{11}$} node[above]
      {\textcolor{red}{$a$}} (x9);
      \path (x7) edge node[left] {$y_{12}$} node[right]
      {\textcolor{red}{$c$}} (x9);
      \path (x7) edge node[left] {$y_{13}$} node[right,pos=.4]
      {\textcolor{red}{$\,b$}} (x10);
      \path (x8) edge node[below] {$y_{14}$} (x10);
      \node (z3) at (1.5,.5) {$z_3$};
      \node (z4) at (1.5,-.5) {$z_4$};
      \node[state] (x11) at (-.8,-2) {$x_{11}$};
      \node[state] (x12) at (.8,-2) {$x_{12}$};
      \node[state] (x13) at (0,-3) {$x_{13}$};
      \path (x0) edge node[right,pos=.8] {$y_{15}$} node[left,pos=.8]
      {\textcolor{red}{$a$}} (x11);
      \path (x0) edge node[left,pos=.8] {$y_{16}$} node[right,pos=.8]
      {\textcolor{red}{$b$}} (x12);
      \path (x11) edge node[left] {$y_{17}$} (x13);
      \path (x12) edge node[right] {$y_{18}$} (x13);
      \node (z5) at (0,-2) {$z_5$};
    \end{scope}
    \begin{scope}[yshift=-28ex]
      \path[fill=black!15] (0,0) to (-1,1) to (-3,1) to (-2,0);
      \path[fill=black!15] (0,0) to (1,1) to (3,1) to (2,0);
      \path[fill=black!15] (0,0) to (-1,-1) to (-3,-1) to (-2,0);
      \path[fill=black!15] (0,0) to (1,-1) to (3,-1) to (2,0);
      \node[state,initial above] (x0) at (0,0) {$x_0'$};
      \node[state] (x1) at (-1,1) {$x_1'$};
      \node[state] (x2) at (-2,0) {$x_2'$};
      \node[state] (x3) at (-1,-1) {$x_3'$};
      \node[state] (x4) at (-3,1) {$x_4'$};
      \node[state] (x5) at (-3,-1) {$x_5'$};
      \path (x0) edge node[left] {$y_1'$} (x1);
      \path (x0) edge node[above] {$y_2'$} (x2);
      \path (x0) edge node[left] {$y_3'$} (x3);
      \path (x1) edge node[below] {$y_4'$} node[above]
      {\textcolor{red}{$b$}} (x4);
      \path (x2) edge node[right] {$\,y_5'$} node[left]
      {\textcolor{red}{$c\,$}} (x4);
      \path (x2) edge node[right] {$y_6'$} node[left]
      {\textcolor{red}{$a\,$}} (x5);
      \path (x3) edge node[below] {$y_7'$} (x5);
      \node (z1) at (-1.5,.5) {$z_1'$};
      \node (z2) at (-1.5,-.5) {$z_2'$};
      \node[state] (x6) at (1,1) {$x_6'$};
      \node[state] (x7) at (2,0) {$x_7'$};
      \node[state] (x8) at (1,-1) {$x_8'$};
      \node[state] (x9) at (3,1) {$x_9'$};
      \node[state] (x10) at (3,-1) {$x_{10}'$};
      \path (x0) edge node[right] {$y_8'$} (x6);
      \path (x0) edge node[above] {$y_9'$} (x7);
      \path (x0) edge node[right] {$y_{10}'$} (x8);
      \path (x6) edge node[below] {$y_{11}'$} node[above]
      {\textcolor{red}{$a$}} (x9);
      \path (x7) edge node[left] {$y_{12}'$} node[right]
      {\textcolor{red}{$c$}} (x9);
      \path (x7) edge node[left] {$y_{13}'$} node[right,pos=.4]
      {\textcolor{red}{$\,b$}} (x10);
      \path (x8) edge node[below] {$y_{14}'$} (x10);
      \node (z3) at (1.5,.5) {$z_3'$};
      \node (z4) at (1.5,-.5) {$z_4'$};
    \end{scope}
  \end{tikzpicture}
  \caption{%
    \label{fi:hdbex2} Two HDA pertaining to Example~\ref{ex:hdbex2}.
    % For the HDA on the
    % left, $\lambda( y_3)= \lambda( y_6)= \lambda( y_9)= \lambda(
    % y_{11})= \lambda( y_{14})= \lambda( y_{15})= \lambda( y_{18})= a$,
    % $\lambda( y_2)= \lambda( y_4)= \lambda( y_7)= \lambda( y_{10})=
    % \lambda( y_{13})= \lambda( y_{16})= \lambda( y_{17})= b$ and
    % $\lambda( y_1)= \lambda( y_5)= \lambda( y_8)= \lambda( y_{12})=
    % c$;
    % on the right, $\lambda'( y_3')= \lambda'( y_6')= \lambda'( y_9')=
    % \lambda'( y_{11}')= \lambda'( y_{14}')= a$, $\lambda'( y_2')=
    % \lambda'( y_4')= \lambda'( y_7')= \lambda'( y_{10}')= \lambda'(
    % y_{13}')= b$ and $\lambda'( y_1)= \lambda'( y_5)= \lambda'( y_8)=
    % \lambda'( y_{12})= c$.
  }
\end{figure}
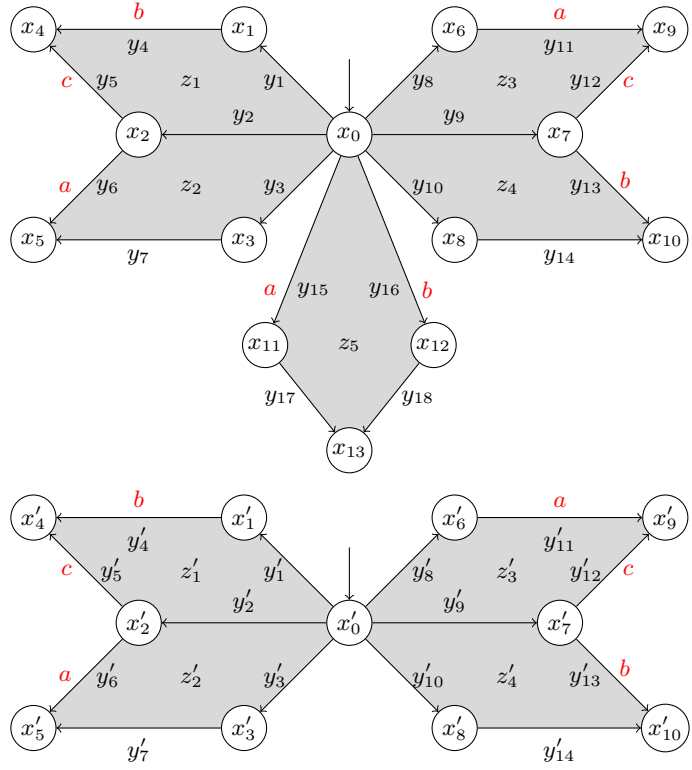

\begin{exa}
  \label{ex:hdbex2}
  We show by a bisimulation-game~\cite{DBLP:conf/banff/Stirling95} type
  argument that the HDA in Fig.~\ref{fi:hdbex2} are not hd-bisimilar.
  Note that in~\cite{DBLP:journals/acta/GlabbeekG01} it is shown that
  these systems are history-preserving bisimilar but not hhp-bisimilar.

  The starting configuration is $( x_0, x_0')$, in which Player~1 (the
  spoiler) plays the $x_0$-extension $y_{16}$.  Player~2 (the
  duplicator) must answer with either $y_2'$ or $y_{10}'$.  Playing
  $y_2'$ is losing, as Player~1 then can play the $y_2'$-extension
  $z_1'$, with label $bc$, which Player~2 cannot duplicate.  Hence
  Player~2 must play $y_{10}'$.  Then Player~1 attacks by extending
  $y_{16}$ with $z_5$, to which Player~2 can only answer $z_4'$.
  Player~1 now retreats to the other lower boundary of $z_5$, $y_{15}$,
  to which Player~2 must answer $y_9'$.  But then Player~1 plays the
  $y_9'$-extension $z_3'$, with label $ac$, which Player~2 cannot
  duplicate.  Hence the game is decided in favor of the spoiler.
\end{exa}

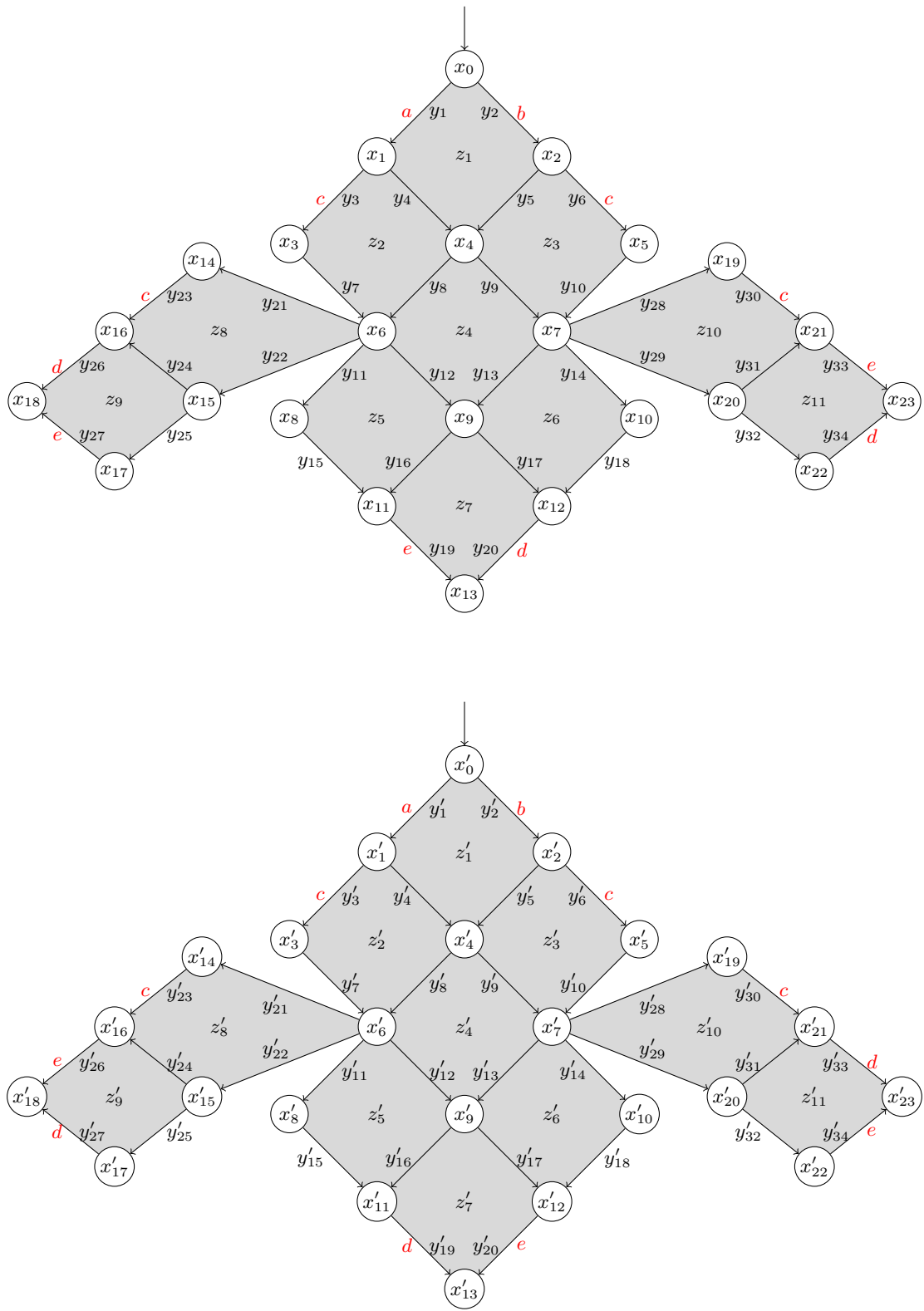
\begin{figure}
  \centering
  \begin{tikzpicture}[auto,scale=1.4,->]
    \tikzstyle{every node}=[font=\footnotesize,initial text=]
    \tikzstyle{every state}=[fill=white,shape=circle,inner
    sep=.3mm,minimum size=6mm]
    \begin{scope}
      \path[fill=black!15] (0,0) to (-2,-2) to (0,-4) to (2,-2);
      \path[fill=black!15] (0,-2) to (-2,-4) to (0,-6) to (2,-4);
      \path[fill=black!15] (-1,-3) to (-3,-2.2) to (-5,-3.8) to
      (-4,-4.6) to (-3,-3.8);
      \path[fill=black!15] (1,-3) to (3,-2.2) to (5,-3.8) to
      (4,-4.6) to (3,-3.8);
      \node[state,initial above] (x0) at (0,0) {$x_0$};
      \node[state] (x1) at (-1,-1) {$x_1$};
      \node[state] (x2) at (1,-1) {$x_2$};
      \node[state] (x3) at (-2,-2) {$x_3$};
      \node[state] (x4) at (0,-2) {$x_4$};
      \node[state] (x5) at (2,-2) {$x_5$};
      \node[state] (x6) at (-1,-3) {$x_6$};
      \node[state] (x7) at (1,-3) {$x_7$};
      \node[state] (x8) at (-2,-4) {$x_8$};
      \node[state] (x9) at (0,-4) {$x_9$};
      \node[state] (x10) at (2,-4) {$x_{10}$};
      \node[state] (x11) at (-1,-5) {$x_{11}$};
      \node[state] (x12) at (1,-5) {$x_{12}$};
      \node[state] (x13) at (0,-6) {$x_{13}$};
      \path (x0) edge node[right] {$y_1$} node[left]
      {\textcolor{red}{$a$}} (x1);
      \path (x0) edge node[left] {$y_2$} node[right]
      {\textcolor{red}{$b$}} (x2);
      \path (x1) edge node[right] {$y_3$} node[left]
      {\textcolor{red}{$c$}} (x3);
      \path (x1) edge node[left] {$y_4$} (x4);
      \path (x2) edge node[right] {$y_5$} (x4);
      \path (x2) edge node[left] {$y_6$} node[right]
      {\textcolor{red}{$c$}} (x5);
      \path (x3) edge node[right] {$y_7$} (x6);
      \path (x4) edge node[right] {$y_8$} (x6);
      \path (x4) edge node[left] {$y_9$} (x7);
      \path (x5) edge node[left] {$y_{10}$} (x7);
      \path (x6) edge node[right] {$y_{11}$} (x8);
      \path (x6) edge node[right] {$y_{12}$} (x9);
      \path (x7) edge node[left] {$y_{13}$} (x9);
      \path (x7) edge node[left] {$y_{14}$} (x10);
      \path (x8) edge node[left] {$y_{15}$} (x11);
      \path (x9) edge node[left] {$y_{16}$} (x11);
      \path (x9) edge node[right] {$y_{17}$} (x12);
      \path (x10) edge node[right] {$y_{18}$} (x12);
      \path (x11) edge node[right] {$y_{19}$} node[left]
      {\textcolor{red}{$e$}} (x13);
      \path (x12) edge node[left] {$y_{20}$} node[right]
      {\textcolor{red}{$d$}} (x13);
      \node[state] (x14) at (-3,-2.2) {$x_{14}$};
      \node[state] (x15) at (-3,-3.8) {$x_{15}$};
      \node[state] (x16) at (-4,-3) {$x_{16}$};
      \node[state] (x17) at (-4,-4.6) {$x_{17}$};
      \node[state] (x18) at (-5,-3.8) {$x_{18}$};
      \path (x6) edge node[below,pos=.6] {$y_{21}$} (x14);
      \path (x6) edge node[above,pos=.6] {$y_{22}$} (x15);
      \path (x14) edge node[right] {$y_{23}$} node[left]
      {\textcolor{red}{$c$}} (x16);
      \path (x15) edge node[right] {$y_{24}$} (x16);
      \path (x15) edge node[right] {$y_{25}$} (x17);
      \path (x16) edge node[right] {$y_{26}$} node[left]
      {\textcolor{red}{$d$}} (x18);
      \path (x17) edge node[right] {$y_{27}$} node[left]
      {\textcolor{red}{$e$}} (x18);
      \node[state] (x19) at (3,-2.2) {$x_{19}$};
      \node[state] (x20) at (3,-3.8) {$x_{20}$};
      \node[state] (x21) at (4,-3) {$x_{21}$};
      \node[state] (x22) at (4,-4.6) {$x_{22}$};
      \node[state] (x23) at (5,-3.8) {$x_{23}$};
      \path (x7) edge node[below,pos=.6] {$y_{28}$} (x19);
      \path (x7) edge node[above,pos=.6] {$y_{29}$} (x20);
      \path (x19) edge node[left] {$y_{30}$} node[right]
      {\textcolor{red}{$c$}} (x21);
      \path (x20) edge node[left] {$y_{31}$} (x21);
      \path (x20) edge node[left] {$y_{32}$} (x22);
      \path (x21) edge node[left] {$y_{33}$} node[right]
      {\textcolor{red}{$e$}} (x23);
      \path (x22) edge node[left] {$y_{34}$} node[right]
      {\textcolor{red}{$d$}} (x23);
      \node at (0,-1) {$z_1$};
      \node at (-1,-2) {$z_2$};
      \node at (1,-2) {$z_3$};
      \node at (0,-3) {$z_4$};
      \node at (-1,-4) {$z_5$};
      \node at (1,-4) {$z_6$};
      \node at (0,-5) {$z_7$};
      \node at (-2.8,-3) {$z_8$};
      \node at (-4,-3.8) {$z_9$};
      \node at (2.8,-3) {$z_{10}$};
      \node at (4,-3.8) {$z_{11}$};
    \end{scope}
    \begin{scope}[yshift=-48ex]
      \path[fill=black!15] (0,0) to (-2,-2) to (0,-4) to (2,-2);
      \path[fill=black!15] (0,-2) to (-2,-4) to (0,-6) to (2,-4);
      \path[fill=black!15] (-1,-3) to (-3,-2.2) to (-5,-3.8) to
      (-4,-4.6) to (-3,-3.8);
      \path[fill=black!15] (1,-3) to (3,-2.2) to (5,-3.8) to
      (4,-4.6) to (3,-3.8);
      \node[state,initial above] (x0) at (0,0) {$x_0'$};
      \node[state] (x1) at (-1,-1) {$x_1'$};
      \node[state] (x2) at (1,-1) {$x_2'$};
      \node[state] (x3) at (-2,-2) {$x_3'$};
      \node[state] (x4) at (0,-2) {$x_4'$};
      \node[state] (x5) at (2,-2) {$x_5'$};
      \node[state] (x6) at (-1,-3) {$x_6'$};
      \node[state] (x7) at (1,-3) {$x_7'$};
      \node[state] (x8) at (-2,-4) {$x_8'$};
      \node[state] (x9) at (0,-4) {$x_9'$};
      \node[state] (x10) at (2,-4) {$x_{10}'$};
      \node[state] (x11) at (-1,-5) {$x_{11}'$};
      \node[state] (x12) at (1,-5) {$x_{12}'$};
      \node[state] (x13) at (0,-6) {$x_{13}'$};
      \path (x0) edge node[right] {$y_1'$} node[left]
      {\textcolor{red}{$a$}} (x1);
      \path (x0) edge node[left] {$y_2'$} node[right]
      {\textcolor{red}{$b$}} (x2);
      \path (x1) edge node[right] {$y_3'$} node[left]
      {\textcolor{red}{$c$}} (x3);
      \path (x1) edge node[left] {$y_4'$} (x4);
      \path (x2) edge node[right] {$y_5'$} (x4);
      \path (x2) edge node[left] {$y_6'$} node[right]
      {\textcolor{red}{$c$}} (x5);
      \path (x3) edge node[right] {$y_7'$} (x6);
      \path (x4) edge node[right] {$y_8'$} (x6);
      \path (x4) edge node[left] {$y_9'$} (x7);
      \path (x5) edge node[left] {$y_{10}'$} (x7);
      \path (x6) edge node[right] {$y_{11}'$} (x8);
      \path (x6) edge node[right] {$y_{12}'$} (x9);
      \path (x7) edge node[left] {$y_{13}'$} (x9);
      \path (x7) edge node[left] {$y_{14}'$} (x10);
      \path (x8) edge node[left] {$y_{15}'$} (x11);
      \path (x9) edge node[left] {$y_{16}'$} (x11);
      \path (x9) edge node[right] {$y_{17}'$} (x12);
      \path (x10) edge node[right] {$y_{18}'$} (x12);
      \path (x11) edge node[right] {$y_{19}'$} node[left]
      {\textcolor{red}{$d$}} (x13);
      \path (x12) edge node[left] {$y_{20}'$} node[right]
      {\textcolor{red}{$e$}} (x13);
      \node[state] (x14) at (-3,-2.2) {$x_{14}'$};
      \node[state] (x15) at (-3,-3.8) {$x_{15}'$};
      \node[state] (x16) at (-4,-3) {$x_{16}'$};
      \node[state] (x17) at (-4,-4.6) {$x_{17}'$};
      \node[state] (x18) at (-5,-3.8) {$x_{18}'$};
      \path (x6) edge node[below,pos=.6] {$y_{21}'$} (x14);
      \path (x6) edge node[above,pos=.6] {$y_{22}'$} (x15);
      \path (x14) edge node[right] {$y_{23}'$} node[left]
      {\textcolor{red}{$c$}} (x16);
      \path (x15) edge node[right] {$y_{24}'$} (x16);
      \path (x15) edge node[right] {$y_{25}'$} (x17);
      \path (x16) edge node[right] {$y_{26}'$} node[left]
      {\textcolor{red}{$e$}} (x18);
      \path (x17) edge node[right] {$y_{27}'$} node[left]
      {\textcolor{red}{$d$}} (x18);
      \node[state] (x19) at (3,-2.2) {$x_{19}'$};
      \node[state] (x20) at (3,-3.8) {$x_{20}'$};
      \node[state] (x21) at (4,-3) {$x_{21}'$};
      \node[state] (x22) at (4,-4.6) {$x_{22}'$};
      \node[state] (x23) at (5,-3.8) {$x_{23}'$};
      \path (x7) edge node[below,pos=.6] {$y_{28}'$} (x19);
      \path (x7) edge node[above,pos=.6] {$y_{29}'$} (x20);
      \path (x19) edge node[left] {$y_{30}'$} node[right]
      {\textcolor{red}{$c$}} (x21);
      \path (x20) edge node[left] {$y_{31}'$} (x21);
      \path (x20) edge node[left] {$y_{32}'$} (x22);
      \path (x21) edge node[left] {$y_{33}'$} node[right]
      {\textcolor{red}{$d$}} (x23);
      \path (x22) edge node[left] {$y_{34}'$} node[right]
      {\textcolor{red}{$e$}} (x23);
      \node at (0,-1) {$z_1'$};
      \node at (-1,-2) {$z_2'$};
      \node at (1,-2) {$z_3'$};
      \node at (0,-3) {$z_4'$};
      \node at (-1,-4) {$z_5'$};
      \node at (1,-4) {$z_6'$};
      \node at (0,-5) {$z_7'$};
      \node at (-2.8,-3) {$z_8'$};
      \node at (-4,-3.8) {$z_9'$};
      \node at (2.8,-3) {$z_{10}'$};
      \node at (4,-3.8) {$z_{11}'$};
    \end{scope}
  \end{tikzpicture}
  \caption{%
    \label{fi:hdbex3}
    Two HDA pertaining to Example~\ref{ex:hdbex3}.}
\end{figure}

\begin{exa}
  \label{ex:hdbex3}
  Again using a hd-bisimulation game, we show that the HDA in
  Fig.~\ref{fi:hdbex3} are not hd-bisimilar.  Note that according
  to~\cite{DBLP:journals/acta/GlabbeekG01},\uli{Where
    in~\cite{DBLP:journals/acta/GlabbeekG01} is this? I can't seem to
    find it.} they are split bisimilar, but not ST-bisimilar.

  From the initial configuration $( x_0, x_0')$ of the game, the spoiler
  plays $y_1$ and then $z_1$, leading to the configuration $( z_1,
  z_1')$.  Playing $y_4$ and then $z_2$, the spoiler forces the
  configuration $( z_2, z_2')$ and, playing $y_8$ and then $z_4$, leads
  the game to the $cc$-labeled configuration $( z_4, z_4')$.  Here the
  spoiler plays $y_{12}$, which the duplicator has to answer by the
  $z_4'$-boundary \emph{in the same direction}, hence $y_{12}'$.  But
  then the spoiler can play the $cd$-labeled $z_5$, to which the
  duplicator has no answer.
\end{exa}

\section{Conclusion}

We have introduced a notion of homotopy bisimilarity for HDA which can
be characterized as an equivalence relation between homotopy classes of
computations, or equivalently by a zig-zag relation between cubes in all
dimensions.  Aside from implying decidability of homotopy bisimilarity
for finite HDA, and together with the results
of~\cite{DBLP:journals/tcs/Glabbeek06}, this confirms that HDA is a
useful formalism for concurrency: not only does it generalize the main
models for concurrency which people have been working with, but it also
is remarkably simple and natural.

One major question which remains is how precisely homotopy bisimilarity
fits into the spectrum of equivalence notions for non-interleaving
models.  We have shown that it is finer than split bisimilarity and
incomparable with history-preserving bisimilarity, but we miss to see
whether homotopy bisimilarity implies ST-bisimilarity and whether it is
implied by hhp-bisimilarity.

% Another important question is how HDA relate to other models for
% concurrency which are not present in the spectrum presented
% in~\cite{DBLP:journals/tcs/Glabbeek06}.  One major such formalism is the
% one of \emph{history-dependent automata} which have been introduced by
% Montanari and Pistore
% in~\cite{DBLP:journals/entcs/MontanariP97,DBLP:conf/stacs/MontanariP97}
% and have recently attracted attention in model
% learning~\cite{DBLP:conf/concur/AartsHV12,DBLP:conf/pts/AartsJU10}.  We
% conjecture that up to history-preserving bisimilarity, HDA are equivalent to
% history-dependent automata.

With regard to the geometric interpretation of HDA as directed
topological spaces, there are two open questions related to the work
laid out in the paper: In~\cite{Fahrenberg05-hda} we show that morphisms
in $\HDA$ are open if and only if their geometric realizations lift
pointed directed paths.  This shows that there are some connections to
weak factorization systems~\cite{AdamekHRT02-weak} here which should be
explored; see~\cite{KurzR05-weak} for a related approach.

In~\cite{Fahrenberg05-thesis} we relate homotopy of cube paths to
directed homotopy of directed paths in the geometric realization.  Based
on this, one should be able to prove that the geometric realization of
the unfolding of a HDA is the same as the universal directed
covering~\cite{FajstrupR08-convenient} of its geometric realization.
%  and
% hence that morphisms in $\HDAh$ are open if and only if their geometric
% realizations lift dihomotopy classes of pointed dipaths.

% The precise relation of our HDA unfolding to the one for Petri
% nets~\cite{DBLP:journals/tcs/NielsenPW81,DBLP:conf/fsttcs/HaymanW08} and
% other models for concurrency should also be worked out.  A starting
% point for this research could be the work on symmetric event structures
% and their relation to presheaf categories
% in~\cite{DBLP:conf/lics/StatonW10}.\uli{FIX AND EXPAND}

\bibliographystyle{plain}
\bibliography{hdabiblong}

\end{document}